\documentclass[a4paper,11pt]{article}
\pdfoutput=1
\usepackage{jheppub}
\usepackage[T1]{fontenc}
\usepackage[english]{babel}
\usepackage{amsthm,amsfonts}
\hyphenchar\font=\string"7F
\usepackage{tensor}
\usepackage[mathstyleoff]{breqn}
\usepackage{slashed}
\usepackage{stmaryrd}
\usepackage{enumerate}
 \usepackage[normalem]{ulem}
\usepackage{tikz}
\usepackage{tensind}
\usepackage[mathscr]{euscript}
\usepackage{mathrsfs}

\usepackage{bbm}

\usepackage{todonotes}

\DeclareMathAlphabet{\mathdsl}{U}{bbm}{m}{sl}
\usepackage{arydshln}
\usepackage{colortbl}

\DeclareMathAlphabet{\mathbbmsl}{U}{bbm}{m}{sl}
\makeatletter
\DeclareFontFamily{OMX}{MnSymbolE}{}
\DeclareSymbolFont{MnLargeSymbols}{OMX}{MnSymbolE}{m}{n}
\SetSymbolFont{MnLargeSymbols}{bold}{OMX}{MnSymbolE}{b}{n}
\DeclareFontShape{OMX}{MnSymbolE}{m}{n}{
    <-6>  MnSymbolE5
   <6-7>  MnSymbolE6
   <7-8>  MnSymbolE7
   <8-9>  MnSymbolE8
   <9-10> MnSymbolE9
  <10-12> MnSymbolE10
  <12->   MnSymbolE12
}{}
\DeclareFontShape{OMX}{MnSymbolE}{b}{n}{
    <-6>  MnSymbolE-Bold5
   <6-7>  MnSymbolE-Bold6
   <7-8>  MnSymbolE-Bold7
   <8-9>  MnSymbolE-Bold8
   <9-10> MnSymbolE-Bold9
  <10-12> MnSymbolE-Bold10
  <12->   MnSymbolE-Bold12
}{}

\let\llangle\@undefined
\let\rrangle\@undefined
\DeclareMathDelimiter{\llangle}{\mathopen}%
                     {MnLargeSymbols}{'164}{MnLargeSymbols}{'164}
\DeclareMathDelimiter{\rrangle}{\mathclose}%
                     {MnLargeSymbols}{'171}{MnLargeSymbols}{'171}
\makeatother

\usepackage{dsfont}
\usepackage{bbm}

\DeclareMathOperator{\Ad}{Ad}

\allowdisplaybreaks
\newcommand{\HIDDEN}[1]{}

\newcommand{\dd}{\mathrm{d}}

\newcommand{\cA}{A}
\newcommand{\cB}{B}
\newcommand{\cC}{C}

\newcommand{\dA}{\mathbb{A}}
\newcommand{\dB}{\mathbb{B}}
\newcommand{\dC}{\mathbb{C}}

\newcommand{\dI}{\mathbb{I}}
\newcommand{\dJ}{\mathbb{J}}

\newcommand{\sa}{\alpha}
\renewcommand{\sb}{\beta}
\renewcommand{\sc}{\gamma}

\newcommand{\ga}{\mathbbm{a}}
\newcommand{\gb}{\mathbbm{b}}
\newcommand{\gc}{\mathbbm{c}}
\newcommand{\gd}{\mathbbm{d}}

\newcommand{\ca}{a}
\newcommand{\cb}{b}
\newcommand{\cc}{c}
\newcommand{\cd}{d}

\newcommand{\CP}[1]{$\mathbb{C}$P$^{#1}$}

\usepackage{tikz}
\usetikzlibrary{matrix,arrows,positioning,shapes,snakes,fadings,decorations.pathmorphing,
decorations.pathreplacing,automata,shadows,decorations.markings,patterns}
\usepackage[thicklines]{cancel}
\usetikzlibrary{calc,shapes.callouts,shapes.arrows}
\newcommand*\circled[1]{\footnotesize\tikz[baseline=(char.base)]{%
            \node[shape=circle,fill=black!20,draw,inner sep=2pt] (char) {#1};}}
            \usepackage{enumitem}

\makeatletter
\def\widebreve{\mathpalette\wide@breve}
\def\wide@breve#1#2{\sbox\z@{$#1#2$}%
     \mathop{\vbox{\m@th\ialign{##\crcr
\kern0.08em\brevefill#1{0.8\wd\z@}\crcr\noalign{\nointerlineskip}%
                    $\hss#1#2\hss$\crcr}}}\limits}
\def\brevefill#1#2{$\m@th\sbox\tw@{$#1($}%
  \hss\resizebox{#2}{\wd\tw@}{\rotatebox[origin=c]{90}{\upshape(}}\hss$}
\makeatletter

\title{\boldmath  Integrable deformation of \CP{n} and\\generalised K\"ahler geometry}
\preprint{MPP-2019-250}

\author[a]{Saskia Demulder,}
\author[b]{Falk Hassler,}
\author[c]{Giacomo Piccinini,}
\author[c,d]{and Daniel C. Thompson}

\emailAdd{sademuld@mpp.mpg.de}
\emailAdd{falk@fhassler.de}
\emailAdd{g.piccinini.987589@swansea.ac.uk}
\emailAdd{D.C.Thompson@Swansea.ac.uk}

\affiliation[a]{Max-Planck-Institut f\"ur Physik, F\"ohringer Ring 6, 80805 M\"unchen, Germany}
\affiliation[b]{George P. \& Cynthia Woods Mitchell Institute for Fundamental Physics and Astronomy,\\ Texas A\&M University, College Station, TX 77843, USA}
\affiliation[c]{Department of Physics, Swansea University, Swansea, SA2 8PP, U.K.}
\affiliation[d]{
Theoretische Natuurkunde, Vrije Universiteit Brussel \& The International Solvay Institutes,\\ B-1050 Brussels, Belgium}

\abstract{We build on the results of \cite{Demulder:2019vvh} 
for generalised frame fields on generalised quotient spaces and study integrable deformations for \CP{n}. In particular we show how, when the target space of the Principal Chiral Model is a complex projective space, a two-parameter deformation can be introduced in principle. The second parameter can however be removed via a diffeomorphism, which we construct explicitly, in accordance with the results stemming from a thorough integrability analysis we carry out. We also elucidate how  the deformed target space can be seen as an instance of generalised K\"ahler, or equivalently bi-Hermitian, geometry. In this respect, we find  the generic form of the pure spinors for \CP{n} and the explicit expression for the generalised K\"ahler potential for $n=1,2$. }

\begin{document}
\maketitle

\section{Introduction} 
Integrable (classical) field theories, a prototypical example being the principal chiral model (PCM), offer a particularly tractable - and yet rich - class of theories. Given the principal chiral model on a group manifold $G$, one might wish to deform it in such a way so as to preserve its integrability whilst generalising the resulting target space geometry, possibly adding a $B$-field contribution. To this end, two models have been particularly successful, namely the $\eta$-deformation of the PCM  (sometimes called Yang-Baxter model)  and the $\lambda$-deformation of WZW CFTs, introduced respectively by Klimcik\cite{Klimcik:2002zj} and Sfetsos\cite{Sfetsos:2013wia}.

Recent work has highlighted a close connection between such integrable models and Poisson-Lie (PL) symmetry.   A non-linear $\sigma$-model is said to have a Poisson-Lie symmetry when the target space manifold, ${\cal M}$, admits the action of a Lie group $G$ (with algebra $\mathfrak{g}$) for which the corresponding world sheet Noether currents  are not conserved (i.e. $G$ is not an isometry group) but obey a modified on-shell  conservation law that is non-commutative with respect to a ``dual'' algebra $\tilde{\mathfrak{g}}$ \cite{Klimcik:1995ux,Klimcik:1996np}.   Such PL models have an elegant  Hamiltonian formalism, now called the ${\mathcal E}$-model \cite{Klimcik:1995dy,Klimcik:1996nq, Klimcik:2015gba}, describing currents on a Drinfel'd double,  i.e. a $2D$-dimensional Lie group, $\mathdsl D$, whose algebra, $\mathfrak{d}$, can be decomposed into two maximally isotropic subalgebras $\mathfrak{g}$ and $\tilde{\mathfrak{g}}$ with respect to an ad-invariant inner product.  The $\eta$-deformation of the PCM on $G$ falls into this class  \cite{Klimcik:2015gba} with the Drinfel'd double given by the complexification of the Lie algebra, $\mathfrak{d}=\mathfrak{g}^{\mathbb{C}}$. Relaxing the requirement that $\mathfrak{d}$ be a Drinfel'd double allows the ${\mathcal E}$-model to also describe the $\lambda$-deformation of WZW CFTs \cite{Orlando:2019his, Thompson:2019ipl}.

The above construction has been generalised in two key ways.  First  one can consider   a multi-parametric integrable deformation of the theories on   group manifolds.  As an example, the so-called bi-Yang-Baxter deformation    \cite{Klimcik:2014bta, Delduc:2015xdm}  augments the deformation parameter, $\eta$, of the  Yang-Baxter model with a second deformation parameter,   $\zeta$, whilst retaining  its (classical) integrability.  Further work in this direction   \cite{Delduc:2017fib, Klimcik:2019kkf, Klimcik:2020fhs} extends this   model including a Wess-Zumino term.

A second generalisation is to consider models defined on spaces other than group manifolds.  This was  initiated in \cite{Klimcik:1996np}, where $\mathcal{E}$-models with a gauge symmetry are shown to  give rise to the ``dressing coset'' construction of Poisson-Lie T-duality. This construction is motivated by the observation that a description without a gauge symmetry is not suitable to capture Abelian and non-Abelian T-duality which act on isometries with fixed point. The additional data this construction requires is an isotropic subgroup $H$ which allows one to realise the target space of the underlying non-linear $\sigma$-model as the double coset ${\cal M}=H \backslash \mathdsl D / \widetilde G$. This setting has come under renewed interest since it includes (the bosonic sector of) integrable  deformations of the $AdS_5 \times S^5$ superstring \cite{Klimcik:2002zj,Delduc:2014kha,Delduc:2013qra,Demulder:2015lva,Hoare:2018ebg,Hoare:2018ngg,Hoare:2015gda,Hoare:2011wr, Hoare:2014pna}. Recently, Poisson-Lie T-duality and variations thereof were studied in the context of non-geometric backgrounds \cite{Osten:2019ayq}.

In this article we will consider integrable deformations on a particular class of coset spaces where $\mathfrak h = \mathrm{Lie}(H)$ is coisotopic. In a first instance, they result in a very natural Ansatz for a deformed Lagrangian containing two deformation parameters. Although such ``double deformations''  have already been considered in a previous work \cite{Sfetsos:2015nya}, their integrability and geometric properties have so far remained quite obscure. In fact, these models come with a precise constraint, detailed in Appendix \ref{app:coisotropy}, which allows for the identification of their target spaces as Poisson homogeneous spaces. In order to address integrability, we shall require these spaces to be additionally Hermitian and symmetric.  As it turns out, complex projective spaces \CP{n} are the natural setting  where to carry out our investigation. They are exemplars of a broader class of algebraic varieties, known as generalised flag manifolds, which have recently benefited, alongside para-complex manifolds,  from a surge of interest  in the physics literature as they materialise in rich examples  \cite{Bykov:2016pfu, Bykov:2016rdv, Bykov:2014efa, Bykov:2015pka,Delduc:2019lpe, Hu:2019zro}.

The key results of this paper include:
\begin{enumerate}[label=\protect\circled{\arabic*}]
 \item  We show how generalised geometry and, in particular, generalised frame fields in the context of $\mathcal{E}$-models are a useful tool to investigate properties of (integrable) $\sigma$-models on both group manifolds and coset spaces. 
 \item If the coset space $G/H$ is an Hermitian and Poisson homogenous space (i.e. $H$ is a coistropic subgroup) we demonstrate that the deformed target space geometry is generalised K\"ahler \cite{Hitchin:2004ut,Gualtieri:2010fd} (equivalently, bi-Hermitian \cite{Gates:1984nk}). As an example, and with future holographic applications in mind, we consider the case of $G/H$= \CP{n} and construct its generalised K\"ahler structure. We show that for $n>1$ the space is parametrised by semi-chiral multiplets plus an additional chiral multiplet when $n$ is odd. If $n=1$ a single chiral multiplet turns out  to be sufficient.  We give an explicit and elegant closed form for the pure spinors corresponding to the generalised complex structures for every $n$ and provide the generalised K\"ahler potential that entirely specifies the geometry for $n=1,2$. 
 \item Writing down a tentative two-parameter deformation when $G/H$ is also a symmetric space we prove, by giving an explicit Lax connection,  the classical integrability of this would-be double deformation. We complete the analysis by finding the twist function describing the Maillet algebra \cite{Maillet:1985ek} of such models. As a byproduct of this analysis, we identify a new tension and deformation parameter making the double deformation equivalent to the single parameter deformation. We confirm this by constructing a diffeomorphism undoing the deformation induced by the second parameter in an appropriate chart. 
\end{enumerate}\

The outline of the remainder of this paper is as follows:  in Section \ref{sec:Worldsheet} we discuss the world sheet of the theories under consideration taking both Lagrangian non-linear $\sigma$-model and Hamiltonian ${\cal E}$-model perspectives. In Section \ref{sec:biYB} we focus our attention to a class of two-parameter integrable Poisson-Lie coset models constructing their Lax representation and demonstrating that they fulfil the requirement of strong integrability. In Section \ref{sec:gkahler} we show that they induce a generalised K\"ahler structure on the target space, culminating in some explicit examples based on \CP{n}.

\section{World sheet perspective} \label{sec:Worldsheet}
The generalised coset construction which we will present in the next section was originally motivated by closed string world sheet theories with Poisson-Lie symmetry. We shall begin in Section \ref{sec:PLsigmamodel} by presenting a conventional Poisson-Lie $\sigma$-model, taking the opportunity to review some known features.  From this perspective one can readily extract conventional target space geometric data even though the underlying algebraic structure is rather obscured. To expose this structure it is helpful to adopt a Hamiltonian approach called the $\mathcal{E}$-model \cite{Klimcik:2015gba} -  introduced in Section~\ref{sec:Emodel} - and construct the corresponding Poisson brackets.

\subsection{Poisson-Lie \texorpdfstring{$\sigma$}{sigma}-models and subgroup invariance}\label{sec:PLsigmamodel}
The Poisson-Lie non-linear $\sigma$-model on a Lie group $G$   is defined by the action\footnote{We use   coordinates $\xi^\pm = \tau \pm \sigma$ such that $\partial_\pm = \frac{1}{2}(\partial_\tau \pm \partial_\sigma)$ with volume element $\dd^2 \sigma = \dd \sigma \dd \tau = \frac12 \dd \xi^+ \dd \xi^-$.}
\begin{equation}\label{eqn:SPLG}
  S_{G} =\frac{1}{\pi} \int_\Sigma \dd^2\sigma \,e_+^\ga \left( E_0^{-1} + \pi_g \right)^{-1}_{\ga\gb} e^\gb_- \, , 
\end{equation} 
in which $e^\ga_\pm$ denote the pull-backs of the left-invariant Maurer-Cartan one-forms $e^\ga= e^\ga{}_i \dd x^i$ onto the world sheet given in terms of a group valued map  $g : \Sigma \rightarrow G$ by  $ e_\pm= g^{-1} \partial_\pm g =   e_\pm^\ga T_\ga$ and $T_\ga$, with $\ga=1,\cdots, D$ and  $D = \mathrm{dim}G$, generate the algebra $\mathfrak g$.  For later use we denote the vector fields dual to these one-forms as $v_\ga= v_\ga{}^i \partial_i$.

The matrix $E_0 = g_0 + b_0$ contains $D^2$ constant entries (which one could promote to fields depending on external spectator coordinates) expressed in terms of a symmetric and a skew-symmetric matrix, $g_0$ and $b_0$, respectively.  $\pi$ is a Poisson-Lie structure compatible with the group multiplication of $G$ \cite{Chari}, i.e.\footnote{For the adjoint action, we adopt the convention $\mathrm{Ad}_g x \equiv g x g^{-1}$. }
\begin{align}\label{eq:Piproperties}
	\pi_{g_1g_2}=\pi_{g_2} + \text{Ad}_{g_2^{-1}}\otimes \text{Ad}_{g_2^{-1}}\pi_{g_1}\,,
\end{align}
thus turning the pair $(G, \pi)$ into a Poisson-Lie group. The Poisson structure induces a Lie algebra structure on the dual vector space $\mathfrak{g}^*$, which we will denote by $\tilde{\mathfrak{g}}$, and we indicate the corresponding structure constants by $\tilde{f}^{\ga\gb}{}_\gc$. The two algebras can be put together so as to form a (classical) Drinfel'd double $\mathfrak{d} = \mathfrak{g} + \tilde{\mathfrak{g}}$, a $2D$-dimensional Lie algebra equipped with an ad-invariant inner product $\llangle \bullet, \bullet \rrangle$ for which $\mathfrak{g}$ and $\tilde{\mathfrak{g}}$ are maximally isotropic subalgebras \cite{Klimcik:1995ux}.

Whilst the action \eqref{eqn:SPLG} is not invariant under global $G$ transformations acting from the left, the properties of $\pi$ ensure that the corresponding would-be-Noether currents $j$, whilst not conserved, still obey a modified non-commutative conservation law \cite{Klimcik:1995ux}
\begin{equation}
\label{eq:conservation}
\dd \star j_\ga = \tilde{f}^{\gb \gc}{}_\ga \star  j_\gb \wedge \star  j_\gc \, . 
\end{equation}
It is this peculiar property that is called a Poisson-Lie symmetry.

By introducing local coordinates $x^i$ on $G$,  one can extract from the action \eqref{eqn:SPLG} a metric $g_{ij}$ and a two-form $b_{ij}$. For later use we also define local coordinates on the double $(x^i,\tilde x_{\tilde \imath})$ on respectively $G$ and $\widetilde G$ in $\mathdsl D=\exp \mathfrak d$ and the corresponding doubled partial derivatives $\partial_I=(\partial^{\tilde\imath},\partial_i)$. In general, $\pi$ will contain rather involved functional dependence on  these local coordinates and the inverted matrix  that defines \eqref{eqn:SPLG} will be complicated, thus rather obscuring the elegance of the action that is indicated by the equation of motion \eqref{eq:conservation}. Instead, a clearer view is obtained by working with the generalised metric which combines the metric and  $B$-field  in a unified object. We will  come back to this point in the next subsection.

An example of a Poisson-Lie symmetric model that will play a particular role in this letter is the well-established   two-parameter integrable model known as the bi-Yang-Baxter theory \cite{Klimcik:2014bta, Klimcik:2008eq}, which is associated to the Drinfel'd double $\mathfrak{d} = \mathfrak{g}^\mathbb{C}$, and for which \begin{equation}\label{eq:E0biYB}
 E_{0}^{-1\, \ga\gb} + \pi^{\ga\gb} =    t\kappa^{\ga\gb} + t \eta R_g^{\ga\gb } + t \zeta R^{\ga\gb }  \, . 
\end{equation}
 Here $\eta$ and $\zeta$ are real valued deformation parameters,  $t$ is the tension for the $\sigma$-model, $g \in G$ is a group element, $\kappa$ is the Killing form, $R^{\ga\gb} = -  R^{\gb\ga}  = \kappa^{\ga\gc} R_{\gc}{}^\gb$ solves the modified classical Yang-Baxter equation and $R_g = \mathrm{Ad}_{g^{-1}} \cdot R \cdot \mathrm{Ad}_{g}$.  As an element of $\mathfrak{g} \wedge \mathfrak{g}$, the Yang-Baxter matrix $R$ is constructed following the Drinfel'd-Jimbo prescription, namely taking the wedge product of (properly normalised) positive and negative roots\footnote{We adopt a Cartan-Chevalley basis $\{H_\lambda, X_\lambda,X_{- \lambda}\}$, where $\lambda \in \Delta_+$ is a positive root, $H_\lambda$ span the Cartan subalgebra and $X_{\pm \lambda}$ are ladder operators. We choose the normalisation, with respect to the Killing form $\langle \cdot, \cdot \rangle$, $\langle X_\lambda, X_{-\lambda}\rangle = \frac{2}{\langle\lambda, \lambda\rangle}$. }
 \begin{equation} \label{eq:drinfeldjimbo}
 R = \frac{1}{2} \sum_{\lambda \in \Delta^+} X_{\lambda} \wedge X_{- \lambda} \, \quad \in \mathfrak{g} \wedge \mathfrak{g} .
 \end{equation} 
  With abuse of notation, we will use the same letter $R$ to indicate the corresponding Lie algebra endomorphism, now operating diagonally on the Cartan-Chevalley generators. The essential properties enjoyed by the latter are that it obeys $R^3= -R$ and projects to zero directions in the Cartan subalgebra of $\mathfrak{g}$. In fact, the $R$-matrix completely captures the Lie algebra structure of the dual bialgebra $\tilde{\mathfrak{g}}$ to $\mathfrak{g}$ inside $\mathfrak d$ through the identity $\tilde{f}^{\ga \gb}{}_\gc =  2 R^{\gd [\ga} f_{\gd \gc}{}^{\gb]}$.

Suppose now  $\mathfrak{g}$ admits a subalgebra $\mathfrak{h}$ such that $\mathfrak{g} = \mathfrak{h} + \mathfrak{m}$ is a symmetric space decomposition. That is, we assume the presence of an involutive Lie algebra automorphism $\sigma$ grading $\mathfrak{g}$, whereby we identify the $\pm 1$  eigenspaces  $\mathfrak{g}^{(0)}$ and $\mathfrak{g}^{(1)}$ with $\mathfrak{h}$ and $\mathfrak{m}$, respectively. We thus split the generators accordingly $T_\ga = \begin{pmatrix} T_{\ca} , & T_{\sa} \end{pmatrix}$  such that $\mathfrak{m}= \textrm{span}( T_{\ca})$ and  $ \mathfrak{h} = \textrm{span}(T_{\sa})$. A natural suggestion for a Poisson-Lie $\sigma$-model on the coset $G/H$ (which we will always take to be reductive) is to restrict the indices in \eqref{eqn:SPLG}  to run over  $\mathfrak{m}$, and so we consider
\begin{equation}\label{eqn:SPLG/H}
  S_{G/H} = \frac{1}{\pi} \int_\Sigma \dd^2 \sigma\, e^{\ca}_+ \left( E_0^{-1} + \pi_g \right)^{-1}_{\ca\cb} e^{\cb}_- \, .
\end{equation}
However, {\em a priori}, the degrees of freedom entering this action still contain those corresponding to the subgroup $H$ and thus, without imposing further constraints, \eqref{eqn:SPLG/H} does not provide a description of the coset $G/H$.  This is remedied by demanding that the action  develops a local gauge symmetry under the action of $H$ from the right which serves to eliminate the unwanted degrees of freedom. A short calculation shows that under an infinitesimal transformation this is the case provided that \cite{Sfetsos:1999zm}
\begin{equation}\label{eqn:constraintE0bis}   
0 = \tilde{f}^{\ca\cb}{}{}_{\sc} + E_0^{-1\, \ca \cd} f_{\sc \cd}{}^{\cb} +f_{\sc \cd}{}^{\ca} E_0^{-1\,\cd \cb}  \, .
\end{equation}
The doubly deformed model we will introduce  will demand us to specialise to the situation where $ \tilde{f}^{\ca \cb}{}{}_{\sc} = 0$.  This can be stated in a more invariant fashion as demanding that $\mathfrak{h}$ is coistropically\footnote{The mathematical terminology used here is detailed in Appendix \ref{app:coisotropy}.} embedded in $\mathfrak g$.  A useful consequence of this condition is that the coset-coset components of the bi-vector evaluated on a subgroup element vanish, i.e. $\pi^{\ca \cb }_h = 0$ for $h\in H$.    As a consequence, under the parametrisation $g = m h$ with $h\in H$, we have from the multiplication law \eqref{eq:Piproperties} that 
\begin{equation}\label{eq:PBforcoset}
  e^{\ca} (g) = (\Ad_{h^{-1}})_{\cb}{}^{\ca} e^{\cb} (m) \, , \quad
 \pi^{\ca \cb}_g = (\Ad_{h^{-1}})_{\cc}{}^{\ca} (\Ad_{h^{-1}})_{\cd}{}^{\cb}  \pi^{\cc \cd}_m  \,.
\end{equation}
This makes it immediately clear that, provided $E_0$ is $H$-invariant,  the action \eqref{eqn:SPLG/H} only depends on the coset representative $m$ in the decomposition $g = m h$.


\subsection{\texorpdfstring{$\mathcal{E}$}{E}-model for group manifolds}\label{sec:Emodel}

In this section we will first define $\cal E$-models for group manifolds and introduce the central concept of generalised frame fields. The latter enables one to access the $\sigma$-model geometry directly from the corresponding $\cal E$-model.

The $\mathcal{E}$-model \cite{Klimcik:1995dy,Klimcik:1996nq, Klimcik:2015gba} is a theory of currents  $ \mathscr{J} = T_\dA  \mathscr{J}^\dA(\sigma) = \mathdsl{g}^{-1} \partial_\sigma \mathdsl{g}$ valued in the loop algebra of $\mathfrak{d} $ which originate from the embedding map  $\mathdsl{g}: \Sigma \rightarrow \mathdsl D$ of the string world sheet into the Drinfel'd double and $T_\dA$ are the $\dim \mathdsl D$ generators of $\mathfrak d$.  The dynamics is generated by the Hamiltonian
\begin{equation} \label{eqn:Hamilton}
  \mathrm{Ham}_{\cal E} = \frac{1}{4\pi} \oint \dd \sigma \llangle  \mathscr{J}, \mathcal{E}  \mathscr{J} \rrangle \, , 
\end{equation}
in which the eponymous operator $\mathcal{E}: \mathfrak{d} \rightarrow \mathfrak{d}$ is an involution, $\mathcal{E}^2= 1$, that is self-adjoint with respect to the inner product  $\llangle\cdot , \mathcal{E}\cdot\rrangle = \llangle \mathcal{E} \cdot , \cdot\rrangle$.  This involution can be specified by $D^2$ parameters which are associated to those of the Poisson-Lie $\sigma$-model by 
\begin{equation}
\mathcal{E}_{\dA}{}^{\dB} = ({\cal H}  \eta^{-1})_{\dA}{}^{\dB}\ , \qquad  \mathcal{H}_{\dA\dB} = \begin{pmatrix}  g_0^{-1} & g_0^{-1} b_0 \\ - b_0 g_0^{-1}    &  g_0- b_0 g_0^{-1} b_0 \end{pmatrix}_{\dA\dB} \, , \qquad \eta_{\dA\dB} = \begin{pmatrix} 0 & &1 \\ 1&  &0 \end{pmatrix}_{\dA\dB} \, , 
\end{equation}
where $g_0$ and $b_0$ are, respectively, a constant symmetric and antisymmetric matrix and $\mathcal H_{\dA\dB}$ is the associated generalised metric.
The Poisson structure of the theory is defined to be a current algebra 
\begin{equation}\label{eqn:emodelpb}
  \{  \mathscr{J}_\dA(\sigma),  \mathscr{J}_\dB(\sigma') \} = 2 \pi  F_{\dA\dB}{}^\dC  \mathscr{J}_\dC (\sigma) \delta( \sigma - \sigma' ) + 2 \pi
    \eta_{\dA\dB} \delta^\prime( \sigma - \sigma' ) \, ,
    \end{equation}
 in which $F_{\dA\dB}{}^\dC$  are the structure constants on $\mathfrak{d}$. Accordingly, we find the equations of motion 
  \begin{equation}\label{eqn:timeevolcurrent}
 \partial_\tau  \mathscr{J} = \{  \mathscr{J}, \mathrm{Ham}_{\cal E} \} = \partial_\sigma \mathcal{E}  \mathscr{J} + [ \mathcal{E}   \mathscr{J},  \mathscr{J} ] \,.
 \end{equation} 
 Taking into account that $\mathcal{E}$ is an involution, we could also  decompose the currents into chiral and anti-chiral parts 
 \begin{align}\label{eqn:chiral}
  \mathscr{J}_\pm = \frac12 ( 1 \pm \mathcal{E})  \mathscr{J}	\, , 
 \end{align}
 and rewrite \eqref{eqn:timeevolcurrent} as
 \begin{equation}\label{eqn:timeevogroup}
 \partial_- \mathscr{J}_+ + \partial_+ \mathscr{J}_- + [ \mathscr{J}_-, \mathscr{J}_+ ] = 0 \,.
 \end{equation}
 Since $\mathcal{E}\mathscr{J}_\pm = \pm \mathscr{J}_\pm$, only half of the components of $\mathscr{J}_+$ and $ \mathscr{J}_-$ are independent.  
 Therefore, although the $\mathcal{E}$-model apparently depends on $2D$ degrees of freedom contained in $\mathdsl{g}$, the first order equations of motion allow half of them (those associated to the subalgebra $\tilde{\mathfrak{g}}$) to be eliminated on-shell to yield second order equations for the rest (those associated to $\mathfrak{g}$).  In this way the dynamics of the theory specified by the $\sigma$-model \eqref{eqn:SPLG} on the target space $\mathcal{M}=\mathdsl D/\widetilde G$ is recovered. The Hamiltonian for a non-linear $\sigma$-model can be expressed in terms of a generalised metric $\mathcal H_{\dI\dJ}$ as
\begin{equation}\label{eq:HamH}
  \mathrm{Ham}_{\cal H} = \frac{1}{4 \pi} \oint \dd \sigma  \begin{pmatrix} 2 \pi  p & \partial_\sigma x\end{pmatrix}^\dI \mathcal{H}_{\dI\dJ} \begin{pmatrix}  2 \pi  p\\ \partial_\sigma x \end{pmatrix}^\dJ 
    \quad \text{with} \quad
  {\cal H}_{\dI\dJ} = \begin{pmatrix} t  g^{-1} & g^{-1} b \\ - b g^{-1}    & t^{-1}( g- b g^{-1} b) \end{pmatrix}_{\dI \dJ}\,,
\end{equation}
in which the canonical momentum is given by $  p_i = (g_{ij} \partial_\tau x^j - b_{ij} \partial_\sigma x^j)/{2 \pi  t}$ and where $(g_{ij},b_{ij})$ is the geometric data entering the $\sigma$-model. Here the indices $\dI$ denote that the fields act on the generalised tangent space of $\mathcal M$ and one has the usual canonical Poisson brackets $\lbrace x^i(\sigma),p_j(\sigma')\rbrace=\delta_j^i\delta(\sigma-\sigma')$.  Given the structure of $\mathrm{Ham}_{\cal H}$, it is natural to introduce some generalised currents $\mathscr{J}_\dI = \left( \partial_\sigma x^i , \, 2 \pi  p_i \right)$ taking values in the generalised tangent space of the target space $\mathcal{M}$. By virtue of the canonical Poisson brackets for $(x,p)$, they can be shown to obey 
\begin{equation}
  \{ \mathscr{J}_\dI(\sigma), \mathscr{J}_\dJ(\sigma') \} = 2 \pi \eta_{\dI \dJ} \delta'(\sigma - \sigma') \, ,
  \end{equation}
  where $\eta$ is again the $O(d,d)$-invariant metric. The same choice of letter for these currents is not accidental as they can be related to the $\cal E$-model currents $\mathscr{J}_\dA$ by introducing  generalised frame fields $E_\dA{}^\dI$ such that
\begin{equation} \label{eqn:Jdefn}
  \mathscr{J}_\dA = E_\dA{}^\dI \mathscr{J}_\dI\,, \quad  E_\dA{}^\dI \eta_{\dI \dJ} E_\dB{}^\dJ = \eta_{\dA \dB} \,, \quad  E_\dA{}^\dI \mathcal{H}_{\dI \dJ} E_\dB{}^\dJ = \mathcal{H}_{\dA \dB}  \, . 
   \end{equation}
Here, and henceforth, $\eta_{\dA \dB}$ and $\eta_{\dI \dJ}$ are used to raise and lower ``doubled'' algebra and target space indices respectively. 
The generalised frame field $E^\dA{}_\dI$ is the generalised geometry analog of  the left  invariant  Maurer-Cartan forms $e^\ga{}_i$ on the target space group manifold ${\cal M}=G$.   For each value of the algebra index $\dA$,  $E_\dA{}^\dI$ defines a generalised vector  - a section of the generalised tangent bundle $ E = T {\cal M} \oplus T^* {\cal M}$ - comprising of a vector field and a one-form   (whose components in a coordinate basis are $E_\dA{}^i$ and $E_\dA{}_i$, respectively).   For a Poisson-Lie $\sigma$-model on a group manifold the generalised frame fields are \cite{Hassler:2017yza}
\begin{equation}\label{eqn:genframeGcompl}
  E_\dA  = \left\{ \begin{array}{l} E^\ga = \pi^{\ga \gb } v_\gb + e^\ga  \\ E_\ga = v_\ga  \end{array} \right. \, ,  
\end{equation}
and obey\footnote{Here we are using the generalised Lie derivative acting on sections of the generalised tangent space. For a pair of generalised vectors $U = u^i \partial_i + \mu_i \dd x^i$ and $V= v^i \partial_i + \nu_i \dd x^i$ the derivative takes the form \begin{equation}\label{eqn:genLieD} {\mathscr L}_{U} V = [u , v] + \left( L_u \nu - \iota_v \dd \mu\right) \,,
\end{equation}
where $L$ is the conventional Lie derivative. }
\begin{equation}\label{eqn:framealggroup}
  \mathscr{L}_{E_\dA} E_\dB = F_{\dA \dB}{}^\dC E_\dC \, .
\end{equation}
This relation is well known in the context of generalised parallellisable spaces and its connection to the $\mathcal{E}$-model was recently studied \cite{Demulder:2018lmj}.

Armed with such a generalised frame field,  the elegant results of Alekseev and Strobl \cite{Alekseev:2004np} show that one can indeed, starting from the canonical Poisson-brackets for $p_i$ and $x_i$ appearing in \eqref{eq:HamH},  derive the Poisson brackets for the currents $ \mathscr{J}_\dA$ as 
\begin{equation}\label{eq:Alekseev}
  \{ \mathscr{J}_\dA(\sigma), \mathscr{J}_\dB(\sigma') \} = 2 \pi  \mathscr{L}_{E_\dA} E_\dB{}^{\dI} E^\dC{}_{\dI} \mathscr{J}_\dC \delta(\sigma - \sigma') + 2 \pi \eta_{\dA \dB} \delta'( \sigma - \sigma' ) \, .
\end{equation}
Upon substitution of \eqref{eqn:framealggroup} into \eqref{eq:Alekseev} one does consistently re-obtain the Poisson-brackets \eqref{eqn:emodelpb}.

 In the remainder of this section, we will take the opportunity to work out and clarify, following \cite{Vicedo:2015pna, Klimcik:2019kkf}, the relation between the $\sigma$-model and the $\mathcal E$-model description. Although we will not aim to encompass all possible $\mathcal E$-models, the construction below will cover most of the relevant models studied to date. 

The equations of motion for the $\sigma$-model on $G$ are well-known to be expressed in terms of some currents $j_\pm \in \mathfrak{g}$; the latter, however, are not in general trivially embedded in their  doubled (anti)chiral counterparts $\mathscr{J}_\pm$. To keep track of this,  we introduce  a surjective  map $S: \mathfrak{g} \times \mathfrak{g} \rightarrow \mathfrak{d}$, called the $S$-map, satisfying the following axioms for all $x,y \in \mathfrak{g}$
\begin{align} \label{eqn:Sconstr1}
  S(x,y)&= S(x,0)  + S(0,y) \, ,   \\
  \label{eqn:Sconstr2}
  S(x,y) &=  0 \quad \text{iff } x=y=0 \, ,  \\
  \label{eqn:Sconstr3}
  S(x,y) &= \mathcal{E} S(x,-y) \, .
\end{align}
Given the splitting \eqref{eqn:chiral}, we are  allowed to identify
\begin{equation}
  \mathscr{J}_+ = S(j_+, 0) \quad \text{and} \quad \mathscr{J}_- = S(0, j_-) \quad \text{with} \quad
  j_\pm \in \Omega^1(\Sigma, \mathfrak{g})\,.
\end{equation}
Particularly interesting are $\mathcal{E}$-models where the time evolution \eqref{eqn:timeevogroup} can be recast as the flatness 
\begin{equation} \label{eqn:classintegr}
  \partial_+ \mathcal{L}_- - \partial_- \mathcal{L}_+ - [\mathcal{L}_- , \mathcal{L}_+] = 0 \, 
\end{equation}
 of a $\mathfrak{g}^\mathbb{C}$-valued Lax connection $\mathcal{L}(z) = \mathcal{L}_+ (z)\dd \xi^+ + \mathcal{L}_-(z) \dd \xi^-$ with spectral parameter $z$. In fact, \eqref{eqn:classintegr} guarantees their classical  weak  integrability.  For example, one can consider a class of models whose equation of motion and Bianchi identity have the form
\begin{equation} \label{eqn:eomBianchigeneral}
\partial_\pm j_\mp +\left [\left( \nu_\pm + \mathcal{O}_\pm  \right)j_\pm, j_\mp \right] = 0  \, , 
\end{equation}
where $\nu_\pm \in \mathbb{C}$ and $\mathcal{O}_\pm$ is an endomorphism of $\mathfrak{g}$. 

Many different models like, for instance, the principal chiral model, the Yang-Baxter model (possibly with a Wess-Zumino term) and the bi-Yang-Baxter model fit into this scenario, upon choosing the appropriate form for $\nu_\pm$ and $\mathcal{O}_\pm$. They all admit the Lax representation  with 
\begin{equation}
\mathcal{L}_\pm = \mathcal{O}_\pm j_\pm + \frac{2 \nu_\pm}{1 \pm z} j_\pm \, .
\end{equation} 
For each of these, in order to recover this result from the $\cal E$-model, a further condition replicating \eqref{eqn:eomBianchigeneral} is needed, namely 
\begin{equation}
[S(0,y), S(x,0)] = S([(\nu_- + \mathcal{O}_-) y,x ], [(\nu_+ + \mathcal{O}_+) x, y]) \, . 
\end{equation}
A similar construction has  already been considered, albeit in a slightly more abstract fashion, in some explicit cases, for instance  for the bi-Yang-Baxter model in \cite{Klimcik:2016rov} where $\mathfrak{d} = \mathfrak{g}^{\mathbb{C}}$.

\subsection{\texorpdfstring{$\mathcal{E}$}{E}-model for coset spaces}\label{sec:Emodelcoset}
On coset spaces  the situation is similar albeit the derivation of the Poisson brackets is more involved. An extensive and general analysis of this setting was recently carried out in \cite{Demulder:2019vvh}; here, we aim at specialising the construction therein to the case at hand. 

We take the generalised coset space\footnote{For the case we will eventually be interested in, i.e. \CP{n}, we will assume that the coisotropic subgroup $H$ sits in $G$. More general situations for the dressing coset construction are addressed in \cite{Demulder:2019vvh}.} $\mathcal{M} =H \backslash \mathdsl D / \widetilde G$ and label its  generators as $T_A =(\widetilde{T}^a, T_a)$, with $\mathfrak{M}= \mathrm{span}(T_A)$, where this partition of generators respectively spans $\mathfrak m^*$ and its dual  $\mathfrak m$. 
In this case we wish to restrict the Hamiltonian to depend only on the currents $ \mathscr{J}= T_{A}  \mathscr{J}^{A}(\sigma)$.   
 To better fit into this formalism, the condition \eqref{eqn:constraintE0bis} can be equivalently rewritten as 
\begin{equation}\label{eqn:Hinvariance}
0 = F_{\alpha A}{}^{C} {\cal H}_{C B} + F_{\alpha B}{}^{C} {\cal H}_{C A} \, , 
\end{equation}
and the Jacobi identity of $\mathfrak{d}$ ensures that  $(T_{\alpha})_{A}{}^{B} = F_{\alpha A}{}^{B}$   generate $\mathfrak{h}$.  Hence, we see that the generalised metric is ad-invariant under $H$-action. As a result the model develops a gauge symmetry and the currents along the subgroup directions become non-dynamical, invoking constraints.  

Just as one can obtain conventional frame fields for a normal coset $G/H$ by reducing and gauge fixing the frames defined on the group $G$, we can obtain  generalised frame fields on the generalised coset by adopting a similar process. A particularly simple example is provided by the case of $H$ being a coisotropic subgroup of a Poisson-Lie group $G$ (recall from the discussion around eq.~\eqref{eqn:constraintE0bis} that this requires $\tilde{f}^{\ca \cb}{}_{\sc} = 0$) since then the Poisson structure on $G$ descends directly to one on $G/H$ \cite{Chari}. This is sometimes called a Bruhat-Poisson structure and denoted by $\pi_{\mathrm{B}}$. Paralleling the discussion for $\cal E$-models on group manifolds, the $\sigma$-model geometry for coset spaces can be obtained by introducing the generalised frame fields given by   \cite{Demulder:2019vvh} 
\begin{equation}\label{eqn:genframecosetcompl}
  E_{A}  = \left\{ \begin{array}{l} E^{\ca} = \pi_{\mathrm{B}}^{\ca\cb } v_{\cb} + e^{\ca}  \\ E_{\ca} = v_{\ca}  \end{array} \right. \, .  
\end{equation}
Here $e^{\ca}$ are the (gauge-fixed) coset components of the left invariant one-forms and $v_{\ca}$ are their Poincar\'e dual vector fields. They arise naturally from the decomposition of the left-invariant Maurer-Cartan form $m^{-1} \dd m = T_{\ca} e^{\ca} + T_{\sa} e^{\sa}$ where $m\in\mathcal{M}$ parameterises the generalised coset space. Similarly to conventional coset spaces, we need to invoke an $\mathfrak h$-valued generalised connection $\Omega^\alpha{}$ (introduced in \cite{Demulder:2019vvh}) entering the generalised frame field on the coset mediating the $H$-compensating transformations and taking care of the gauge invariance,
\begin{equation}\label{eqn:OmegaalphaB}
  \Omega^\alpha{} = e^{\sa} + \pi^{\sa \cb} v_{\cb} \,.
\end{equation}
 In order to find the frame algebra obeyed by the generalised frame field, we make use of the machinery in \cite{Demulder:2019vvh} to obtain 
\begin{equation}\label{eqn:framealgcoset} 
  \begin{aligned}
    \mathscr{L}_{E_A} E_B &= F_{AB}{}^C E_C + 2 \Omega^\alpha{}_{[A} F_{\alpha B]}{}^C E_C + F_{\alpha AB} \Omega^{\alpha} \, , \\
    \mathscr{L}_{\Omega^{\alpha}{}} E_B &=  (\Omega^{\alpha\beta} F_{\beta B C} + F^\alpha{}_{BC}) E^C + F^{\alpha}{}_{\beta\gamma} \Omega^{\beta}{}_B \Omega^\gamma \, , 
  \end{aligned}
\end{equation} 
where  $\Omega^\alpha=\Omega^{\alpha B}E_B$, so that  
\begin{equation}\label{eqn:emodelpbcoset}
  \{  \mathscr{J}_{A}(\sigma),  \mathscr{J}_{B}(\sigma') \} = 2 \pi  \mathcal{F}_{A B C}(\sigma)  \mathscr{J}^{C}(\sigma) \delta( \sigma - \sigma' ) + 2 \pi \eta_{A B} \delta^\prime( \sigma - \sigma' ) \, .
\end{equation}
Here the currents $\mathscr{J}_A$ are defined in the same spirit  as in \eqref{eqn:Jdefn}, where we now restrict  indices to run over generalised coset generators, $ \dA \rightarrow A$, and coordinates, $\dI \rightarrow I$.
In contrast to the case of a group manifold, the $\mathcal F_{A B C}$ appearing in \eqref{eqn:emodelpbcoset} are no longer constants but instead given by
\begin{align}
	\mathcal F_{A B C}(\sigma)= F_{A B C}+3\Omega_{[A B C]}(\sigma)\,,
\end{align}
where $\Omega_{AB}{}^C = \Omega^\delta{}_
A F_{\delta B}{}^C$.
Enforcing the corresponding Jacobiator one obtains the consistency requirement 
\begin{equation}\label{eqn:BPconstr}
  D_{  [A}  \mathcal F_{B C D]} \overset{!}{=}  \frac34
  \mathcal   F_{[A B}{}^{E} \mathcal   F_{ E CD]}  \, , 
\end{equation}  
where we have introduced the derivative operator $D_A=E_A{}^I\partial_I$ obtained by contracting the doubled partial derivative with respect to the generalised coset coordinates  with the frame fields restricted as above. Due to the properties of the compensator $\Omega$ and the generalised frame field this condition is satisfied by construction \cite{Demulder:2019vvh}. 
To completely characterise the system we shall also require the Poisson brackets for the compensator \begin{equation}\label{eqn:PBJomega}
  \{  \mathscr{J}_{A}(\sigma), \Omega^{\sb}{}_{C}(\sigma') \} = 2 \pi D_{A} \Omega^{\sb}{}_{C} \delta(\sigma - \sigma')\,, \quad   \{ \Omega^{\sa}{}_{A}(\sigma), \Omega^{\sb}{}_{B}(\sigma') \} = 0 \, . 
\end{equation}
The Jacobiators involving the combinations $(\mathscr{J}  , \mathscr{J} , \Omega)$, $(\mathscr{J}  , \Omega , \Omega)$ and $(\Omega, \Omega, \Omega)$ vanish identically as well.

The equations of motion  can be cast in a form  similar  to \eqref{eqn:timeevogroup} by introducing the quantity
$\mathscr{B}_\pm = T_{\sa} \mathscr{B}^{\sa}_\pm = T_{\sa} \Omega^{\sa}{}_{B} \mathscr{J}_\pm^{B} $
to find equations for the currents $ \mathscr{J}_{\cA}(\sigma)$
\begin{equation}\label{eqn:timeevoJcoset}
\partial_-  \mathscr{J}_+  + \partial_+ \mathscr{J}_- + [  \mathscr{J}_-,  \mathscr{J}_+ ] + [\mathscr{B}_-,  \mathscr{J}_+] + [\mathscr{B}_+,  \mathscr{J}_-] = 0\,.
\end{equation}
Note that the two objects  $\mathscr{J}_\pm  =  \mathscr{J}_\pm^{A } T_{A}$ are defined as in eq.~\eqref{eqn:chiral}. While the equations of motion for $\mathscr{B}_\pm$ read
\begin{equation}\label{eqn:eomB}
\partial_+ \mathscr{B}_- - \partial_- \mathscr{B}_+ + [\mathscr{B}_+,\mathscr{B}_-] +  \mathcal{P}_{\mathfrak{h}} [\mathscr{J}_+,\mathscr{J}_-] = 0  \, , 
\end{equation}
where $\mathcal{P}_{\mathfrak{h}}$ is a projector from $\mathfrak{d}$ to the isotropic subgroup $\mathfrak{h}$. 

Let us now construct the $S$-map explored above for coset $\mathcal E$-models when $H$ is a coisotropic subgroup and $G/H$  a symmetric space. This latter assumption is sufficient to prove that the dual algebra $\tilde{ \mathfrak{g}}$ equipped with the Lie bracket induced by the $R$-matrix gives rise to the symmetric space $\widetilde{M}=\widetilde{G}/H^*$, $\mathfrak{h}^* = \mathrm{Lie}(H^*)$  being  the vector space dual to $\mathfrak{h}$ (see Appendix \ref{app:coisotropy} for a concise proof).  As a consequence,  the structure coefficients $F_{A B C}$ vanish  and with them so does the projection along $\mathfrak{M}$ of the bracket $[\mathscr{J}_-, \mathscr{J}_+]$. 
In such a scenario we wish to express the currents $\mathscr{J}_\pm$ valued in $\mathfrak M$ in terms of the $\mathfrak m$-valued currents $j_\pm$ using the $S$-map as before. To this end, suppose that $S : \mathfrak{g} \times \mathfrak{g} \rightarrow \mathfrak{d}$ satisfies the two  constraints  \eqref{eqn:Sconstr1} and \eqref{eqn:Sconstr2}. Assume now it admits a restriction to the coset $\mathfrak m$  such that  its image lies in $\mathfrak{M}$, $\mathrm{Im} S (\mathfrak{m},\mathfrak{m}) \subset \mathfrak M$. The constraint \eqref{eqn:Sconstr3} is preserved, provided that we now interpret, using the coisotropy property of $\mathfrak h$, $\mathcal{E}$ as the restriction of the generalised metric to the generalised coset, i.e. $\mathcal{E}: \mathfrak{M} \rightarrow \mathfrak{M}$,  and take $x,y \in \mathfrak m$. In addition, we impose that for all $ x,y \in \mathfrak m$
\begin{equation} \begin{gathered}
\label{eqn:SCoset}
[x,y]  = \mathcal{P}_{\mathfrak{h}} [S(x,0), S(0,y)]   \, ,  \qquad 
  [T_{\sa}, S(x, y)]= S( [T_{\sa}, x ], [T_{\sa}, y] ) \, .
\end{gathered} \end{equation}
The field equation \eqref{eqn:timeevoJcoset} is then equivalent to
\begin{equation}\label{eqn:symspaceintegrable1}
  \partial_- j_+ + [ \mathscr{B}_-, j_+ ] = 0 \, , \qquad  \partial_+ j_- + [ \mathscr{B}_+, j_- ] = 0 \, ,
\end{equation}
while the Bianchi identity reads
\begin{equation}\label{eqn:symspaceintegrable2}
  \partial_+ \mathscr{B}_- - \partial_- \mathscr{B}_+ + [\mathscr{B}_+,\mathscr{B}_-] + [j_+,j_-] = 0\,.
\end{equation}
Such theories are classically integrable and their dynamics is  encoded in the flatness of the Lax connection 
\begin{equation}\label{eqn:laxsymspace}
  \mathcal{L}(z) = \mathscr{B}_+ \dd \xi^+ + \mathscr{B}_- \dd \xi^- + z j_+ \dd \xi^+ + \frac{1}{z} j_- \dd \xi^- \ , \quad \mathscr{B}_\pm \in \Omega^1(\Sigma, \mathfrak{h}) \, , ~~ j_\pm \in \Omega^1(\Sigma, \mathfrak{m}) \, . 
\end{equation}
To give a flavour of $S$ in a case relevant to the discussion ahead, the map for a Yang-Baxter deformation of \CP{n} turns out to be\footnote{We assume that $R$ here is a restriction of the Yang-Baxter matrix to the sole coset directions, i.e. $R : \mathfrak{m} \rightarrow \mathfrak{m}$. It acts on an element $x \in \mathfrak{m}$ via $R(x) = x^b R_{b}{}^a T_a$. Analogously, $\kappa$ is the restriction of the Killing form to the generators of $\mathfrak{m}$.}, for every $x, y \in \mathfrak{m}$, 
\begin{equation}
S(x,y) = \frac{1}{\sqrt{1+ \eta^2}}  \begin{pmatrix}
x+y +  \eta R(x-y) \\
t^{-1} \kappa (x-y)
\end{pmatrix} \,  ,
\end{equation}
where $\eta$ is the deformation parameter, $t$ the tension of the $\sigma$-model  and the upper and lower components refer to $\mathfrak{m}$ and $\tilde{\mathfrak{m}}$, respectively, where we use $\tilde{\mathfrak{m}}$ to denote the linear space such that $\exp(\tilde{\mathfrak{m}}) \cong \widetilde{M} = \widetilde{G}/H^*$.


\section{Integrable deformations of \texorpdfstring{\CP{n}}{CPn}} \label{sec:biYB}
In this section we shall put the general machinery presented  thus far to use. We start by considering a  double deformation already introduced in \cite{Sfetsos:2015nya} consisting of the deformation parameters $\eta$ and $\zeta$. However, as it transpires from the study of its integrability structure, the second parameter $\zeta$ does not play a role, for it can be reabsorbed in new effective tension and deformation parameter. At the geometrical level  we can equivalently construct a diffeomorphism undoing the deformation introduced by $\zeta$.
 Remarkably though, the deformed geometry of \CP{n} spaces can be  elegantly described in terms of generalised K\"ahler geometry. 
 All the geometrical aspects will be the subject of Section \ref{sec:gkahler}, while we devote the present one to the study of its integrability. 

\subsection{\texorpdfstring{$\sigma$}{sigma}-model description}
Let us first return to the constraint of $H$-invariance required on $ E_0^{-1}|_{\mathfrak m}$ in eq.~\eqref{eqn:constraintE0bis} (or equivalently eq.~\eqref{eqn:Hinvariance} on the generalised metric).  We wish to consider specific choices of $(\mathfrak{d}, \mathfrak{g}, \mathfrak{h})$ for which the admissible  ${\cal H}_{\cA \cB}$ allows for multiple free parameters.  A model with such features was already identified in \cite{Sfetsos:2015nya}. There it was shown that $H$-invariance of its generalised metric  implied the constraint
\begin{equation} \label{eq:coisotropycond}
  \tilde{f}^{\ca \cb}{}_{\sc} =0 \Leftrightarrow \left([x, Ry] + [Rx, y] \right) |_\mathfrak{h} = 0 \quad \forall x, y \in \mathfrak{m} \, .
\end{equation}
When this condition holds,  $\mathfrak{h}$  is said to be a coisotropically embedded subalgebra, as introduced above in Section \ref{sec:PLsigmamodel}, and the Poisson structure on $G$ descends to one on $G/H$ making it a Poisson homogenous space\footnote{ Notice that this condition holds for $\mathfrak{g}=\mathfrak{so}(n+1)$ and $\mathfrak{h}= \mathfrak{so}(n)$ so that one could think of applying the same construction to $S^n$ \cite{panyushev2019poissoncommutative}. However, in this case the restriction $R|_{\mathfrak{m}}$ vanishes, making the inclusion of the second parameter trivial from the outset.}. 
	
In what follows it will be crucial to refine, motivated by the goal of integrability, the set of coset spaces to Hermitian symmetric spaces, implying that $R_{\ca}{}^{\cc}R_{\cc}{}^{\cb}= -\delta_{\ca}{}^{\cb}$ i.e. the $R$-matrix descends to a complex structure (see Appendix \ref{app:coisotropy} for a more precise treatment). 
The canonical examples of such a set-up are the complex Grassmannians for which $G= SU(n+m)$ and $H= S(U(n) \times U(m))$, of which  \CP{n} is a particularly interesting subcase\footnote{Regarding \CP{n} $\sigma$- models  we are making statements only about classical integrability; one should anticipate that quantum integrability necessitates the inclusion of appropriate fermionic content \cite{Abdalla:1980jt,Abdalla:1981yt}. Fortunately our results point to how this should be done since they will be compatible with $\mathcal{N}=2$ world sheet supersymmetry.  }.  

Thus, assuming that $G/H$ is an Hermitian symmetric space with $H$ a coisotropic subgroup, the $\sigma$-model 
reads
\begin{equation} \label{eq:biYBcoset}
  S = \frac{1}{\pi  t} \int_\Sigma \dd^2 \sigma \langle \mathcal{P}_1 e_+(m), \frac{1}{1 - \eta R_m \mathcal{P}_1 - \zeta R \mathcal{P}_1} \mathcal{P}_1  e_-(m) \rangle \: ,
\end{equation}
in which $R_m = \mathrm{Ad}_{m^{-1}} \cdot R \cdot \mathrm{Ad}_{m}$ and where, to match notation elsewhere, we introduce  projectors $\mathcal{P}_i$ onto $\mathfrak{g}^{(i)}$ in the symmetric space decomposition and let $\kappa_{\ca\cb} = \langle T_\ca , T_\cb \rangle $. Because of gauge invariance, the action \eqref{eq:biYBcoset} depends only on the coset representative $m$ in the parametrisation $g= m h$ with $h \in H$.

\subsection{Weak and strong integrability}
Proving that a model is (classically) integrable in the weak sense amounts to find a Lax connection whose flatness implies the equations of motion and Bianchi identity. To this end we introduce the currents\footnote{To make contact with the literature we adopt the notation in \cite{Delduc:2013fga}.}
 	\begin{equation} \label{eq:Bpm}
  B_\pm = \frac{1}{1 \pm \eta R_m \mathcal{P}_1 \pm \zeta R \mathcal{P}_1} e_\pm(m)\, , \qquad j_\pm = k B_\pm^{(1)} \pm \frac{\zeta}{k} \mathcal{P}_1 R B_\pm^{(1)} \,  ,
	\end{equation}
	where $k$ is the combination of parameters solving $k^4 - (1+ \eta^2- \zeta^2) k^2 - \zeta^2=0$. 
The Lax connection can then be shown to be 
	\begin{equation} \label{eqn:Lax}
		\mathcal{L}_\pm = B_\pm^{(0)} + z^{\pm 1} j_\pm \: .
	\end{equation}
	As for strong integrability, a sufficient condition  is to ascertain that the Lax matrix $L(z)=\mathcal{L}_+(z) - \mathcal{L}_-(z)$ obeys the Maillet algebra \cite{MAILLET1986401,Maillet:1985ek}
	\begin{equation} \label{eqn:MailletAlgebra}
	\begin{split}
	\{{L}_1 (\sigma, z), {L}_2 (\sigma', w)\}=& [r_{12} (z,w), {L}_1 (\sigma ,z) + {L}_2 (\sigma, w)] \delta (\sigma- \sigma')  \\
	+& [s_{12}(z,w), {L}_1 (\sigma, z) - {L}_2 (\sigma, w)] \delta(\sigma - \sigma') - 2  s_{12}(z,w) \delta'(\sigma - \sigma') \, , 
	\end{split}  
	\end{equation} 
	for a specific choice of the matrices $r$ and $s$. These two are, respectively, the skew-symmetric and symmetric part of a solution $\mathcal{R}$ of the mCYBE for the loop algebra of $\mathfrak g$. Of particular interest are the models for which $\mathcal{R}$ is fully specified by a model independent $\mathcal{R}_0$, usually expressed in terms of the graded projections of the Casimir operator\footnote{The superscripts refer to the corresponding subspace in the symmetric space decomposition, $i,j=0,1$. }   $\mathcal{C}^{(ij)}$ of $\mathfrak{g}$, multiplied by a scalar function of the spectral parameter $\varphi(z)$, known as the twist function \cite{Vicedo:2010qd}. In this case and when considering coset spaces, $r$ and $s$ assume the form 
\begin{align}
r_{12}(z,w) &= - \frac{1}{2} \frac{1}{z^2 -w^2} \left[ (z \varphi^{-1}(z) + w \varphi^{-1}(w)) \mathcal{C}^{(00)}_{12} + (w \varphi^{-1}(z) + z \varphi^{-1}(w)) \mathcal{C}^{(11)}_{12}  \right] \,  , \\
s_{12}(z,w) &= + \frac{1}{2} \frac{1}{z^2 -w^2} \left[ (z \varphi^{-1}(z) - w \varphi^{-1}(w)) \mathcal{C}^{(00)}_{12} + (w \varphi^{-1}(z) - z \varphi^{-1}(w)) \mathcal{C}^{(11)}_{12}  \right] \,  . 
\end{align}
	The $\mathcal{E}$-model formalism  facilitates the analysis; we first introduce the Lie algebra valued quantities
	\begin{equation}
  X^{\ga}= e^{\mathbbm{a}\mathbbm{i}}(g) p_ \mathbbm{i}  \, , \qquad Y^\mathbbm{a} = \frac{1}{2 \pi t} e_\sigma^\mathbbm{a}(g) - X^\mathbbm{b} (\eta R_g + \zeta R)_{\mathbbm{b}}{}^\mathbbm{a} \,  ,  
	\end{equation}
	where we take $g = m h$ and $p$  the canonical momentum.  The advantage of these definitions is that the Lax matrix can be written as a function of $X$ and $Y$. In turn, these are conveniently embedded into a generalised vector via $\mathscr{Z}_{\dA} = 2 \pi (t Y^\ga, X_\ga)$. The crucial observation is that the latter is  related to the currents $\mathscr{J}_\mathbb{A}$ introduced in Section \ref{sec:Emodel} by  a constant  $O(d,d)$ transformation $\beta$ 
	\begin{equation}
	\mathscr{Z}_\dA =  \beta_\dA{}^\dB \mathscr{J}_\dB \, \qquad \text{with} \qquad \beta = \begin{pmatrix}
	1 &t  (\eta+ \zeta) \kappa^{-1}R \\ 0 & 1
	\end{pmatrix} \,  .
	\end{equation}
	As a result, the Poisson brackets for $\mathscr{Z}$, and thus the ones for $L(z)$, can be inferred from the ones for $\mathscr{J}$ already given in \eqref{eqn:emodelpb}. Eventually, the Poisson brackets for the Lax matrix can be used to check if the Maillet algebra \eqref{eqn:MailletAlgebra} is satisfied. As it turns out, our model fits into the class of models with twist function, where we find
	\begin{equation}
  \varphi(z) = \frac{k^2}{t\pi (k^2 + \zeta^2)} \frac{z}{(z^2-1)^2 + \frac{k^2 (k^2-1)}{k^2 + \zeta^2} (z^2+1)^2 } \,  .
	\end{equation}
	Upon identifying  new deformation parameter  and tension according to
	\begin{equation} \label{eq:tildet}
  \tilde \eta^2 = \frac{k^2 (k^2-1)}{k^2 + \zeta^2} \, , \qquad \tilde t =  t \frac{k^2 + \zeta^2}{k^2}  \,  ,
	\end{equation}
	one can see that the twist function above coincides with the one for the single Yang-Baxter deformation already present in the literature \cite{Delduc:2013fga}.  It is tempting to infer, judging from this analysis, that the second parameter $\zeta$ is not affecting the model. Nevertheless, given the existence of transformations not affecting the twist function \cite{Vicedo:2015pna} and yet yielding to a non trivially deformed target space, a  more detailed geometric check is needed. We will explicitly construct the diffeomorphism removing $\zeta$ from the deformed metric in the next section.  

\subsection{Renormalisation group flow }
A useful cross-check of the above result comes from  analysing the renormalisation group flow for the effective parameters $\tilde{\eta}$ and $\tilde{t}$. Using the doubled formalism  introduced in Section~\ref{sec:Emodelcoset}, the renormalisation group flow can be written as
\begin{equation}
  \frac{\dd \mathcal{H}_{\cA\cB}}{\dd \log\mu} \propto \widehat{\mathcal{R}}_{\cA\cB}\, ,
\end{equation}
where $\widehat{\mathcal{R}}_{\cA \cB}$ denotes a projection of the generalised Ricci tensor for a generalised symmetric space whose form can be found in eq. (3.36) in \cite{Demulder:2019vvh}. An equivalent form of the flow equation, which is the one we shall eventually make use of, is given by
\begin{equation}
  \frac{\dd \mathcal{H}}{\dd \log\mu} \mathcal{H}^{-1} \propto [ \mathdsl{O}, \mathcal{E} ] 
    \quad \text{with} \quad   \mathdsl{O} = \begin{pmatrix}
     \eta t  R^t & 0\\
    \kappa & \eta t R 
  \end{pmatrix} \, .
\end{equation}

Of the three parameters $\eta$, $\zeta$ and $t$ entering eq.~\eqref{eq:biYBcoset}  there are two RG invariants which can be chosen to be  $\eta t$ and $\frac{1+ \zeta^2 +\eta^2}{\eta \zeta}$.   The remaining non-trivial flow equation is obtained as\footnote{For the case of $S^2=$ \CP{1} this coincides with  the result   obtained through geometric methods in \cite{Sfetsos:2015nya}. A second verification of this result can be obtained by considering the RG equations obtained for PL-models on a group space in \cite{Sfetsos:2009vt} and applying the limit procedure described in  \cite{Sfetsos:1999zm}. } 
	\begin{equation} \label{eq:flow}
	\frac{\dd \eta}{\dd \log\mu} \propto {\eta t} (1- \zeta^2 + \eta^2)\, . 
	\end{equation}
	One can prove that $\tilde{\eta} \text{ and } \tilde{t}$,  as defined in eq.  \eqref{eq:tildet},  obey the same flow eq. \eqref{eq:flow} (upon setting $\zeta=0$). In particular, the fact that $\eta t = \tilde{\eta} \tilde{t}$ implies that the latter  is again an invariant.

\section{Generalised K\"ahler structure of deformed \texorpdfstring{\CP{n}}{CPn}}\label{sec:gkahler}
In this section we show how the backgrounds presented in the previous section are, notwithstanding their apparent complexity, exemplars of generalised K\"ahler manifolds. 

 \subsection{Target space geometry}
       
We now turn to the curved target space geometry.  One can extract a conventional metric $g$ and NS two-form field $b$ either from directly evaluating the $\sigma$-model action of eq.~\eqref{eq:biYBcoset} or equivalently by dressing the generalised metric
\begin{equation}\label{eq:genmet1}
  \mathcal{H}_{\cA\cB} = \begin{pmatrix}
  (1 + (\eta + \zeta)^2) t \kappa^{-1} &  ( \eta + \zeta ) R^t \\
     (\eta + \zeta)  R &  t^{-1} \kappa
  \end{pmatrix}{} 
\end{equation}
with the generalised frame fields which we recall are given by\footnote{With respect to eq. \eqref{eqn:genframecosetcompl} we have introduced the factor $\eta t$ needed to ensure the equivalence between the Poisson-Lie model \eqref{eqn:SPLG/H} and the action \eqref{eq:biYBcoset}. }
\begin{equation}\label{eq:genFFcoset}
  E_{\cA}  = \begin{pmatrix} e^{\ca} && \eta t \pi_{\mathrm{B}}^{\ca \cb} {e^{-t}}_{\cb} \\ 0 && e^{-t}_{\ca} \end{pmatrix}= \begin{pmatrix} \delta^{\ca}_{\cb} && \eta t \pi_{\mathrm{B}}^{\ca \cb} \\ 0 && \delta^{\cb}_{\ca} \end{pmatrix} \begin{pmatrix} e^{\cb} && 0 \\ 0 && e^{-t}_{\cb} \end{pmatrix} \, , 
 \end{equation}      
in which we have introduced a coset representative $ m$ with $m^{-1} \dd m = e^{\ca} T_{\ca} + e^{\sa } T_{\sa}$. Here $\pi_\mathrm{B}$ is the Bruhat-Poisson structure on the Poisson homogenous space  $G/H$, which we can express as
\begin{equation}
\pi_{\mathrm{B}}^{\ca \cb} =  (R_m - R)^{\ca \cb}  \, ,
\end{equation} 
where we recall that  $R_m = \mathrm{Ad}_{m^{-1}} \cdot R \cdot \mathrm{Ad}_{m}$. 
By assumption the target space is in addition equipped with a complex structure, an Hermitian metric (e.g.  the Fubini-Study metric in the case of \CP{n}), and a K\"ahler two-form that we denote respectively by 
\begin{equation}
J = e^{\ca} R_{\ca}{}^{\cb} e^{-t}_{\cb} \ , \quad  G =  e^{\ca} \kappa_{\ca \cb} e^{\cb} \ , \quad \omega = JG  \,  . 
\end{equation} 
It will be convenient to observe that in this set up we can introduce a more general bi-vector consisting of the linear combination of the Bruhat-Poisson structure and the inverse of the K\"ahler-Fubini  structure, forming a so-called Poisson pencil, 
\begin{equation} \label{eqn:PP}
\pi_\tau = \pi_{\mathrm{B}} - \tau \, \omega^{-1}   \qquad   \tau \in \mathbb{R} \, , 
\end{equation} 
which, for \CP{n} at least\footnote{In general $\pi_\tau$ does not obey the Schouten identity. However, as proven in \cite{khoroshkin1993},  it does on any Hermitian symmetric space (see Appendix \ref{app:coisotropy} for a thorough discussion).}, obeys the Schouten identity and defines a genuine Poisson structure.  After some manipulations one finds that the deformed geometry encapsulated by $g \text{ and }b$ can be expressed in terms of $G$ and $\pi_\tau$ as
	\begin{equation} \label{eq:biYBgeom}
		\begin{gathered}
		g^{-1}= G^{-1} - \eta^2 \pi_{\tau} G \pi_{\tau}  \, , \qquad
		bg^{-1} =-  \eta \, G  \pi_{\tau}   \, ,
		\end{gathered}
	\end{equation}
	in which the Poisson pencil parameter is fixed to 
	$
	\tau = 1 + \frac{\zeta}{\eta}\, . 
	$
	Let us emphasise that, despite the elegant form of  \eqref{eq:biYBgeom}, in terms of explicit coordinate expressions these become rather intractable.

 \subsection{From  double to  single parameter deformation} \label{sec:trivial}
Let us start by recalling the Fubini-Study metric on  \CP{n}
\begin{equation}
\begin{aligned}
\dd s^2   =& \frac{\dd z_i \dd \bar{z}_i }{1 + |z|^2 } - \frac{z_i \bar{z}_j \dd z_j \dd \bar{z}_i }{(1 + |z|^2)^2  } \, , 
\end{aligned}
\end{equation}
in which the coordinates $z_i$, which we will similarly refer to as Fubini-Study (FS) coordinates, are holomorphic with respect to a complex structure $J = i  \dd z_i \otimes \partial_{z_i} -i  \dd \bar{z}_i \otimes \partial_{\bar{z}_i}$ and $|z|^2= z_i \bar{z}_i$. More precisely, we put ourselves in the patch, often called the largest Bruhat cell, where the first homogeneous coordinate is not vanishing. 
Let us now introduce coordinates $(x_i, \phi_i )$ better adapted to the deformed geometry;  they are defined by 
\begin{equation} \label{eqn:xphi}
  z_i  = \left( \frac{x_i}{1 - X} \right)^{1/2} e^{i \phi_i} \,  , \quad \text{with} \quad 0 \leq x_i < 1- \sum_{k=1}^{i-1}x_k \quad \text{and} \quad 0 \leq \phi_i < 2 \pi \, , 
\end{equation}
 where $X = \sum_{i} x_i$. With respect to these, the Poisson-Bruhat structure reduces to 
\begin{equation} \label{eqn:piB}
\pi_{\mathrm{B} } =  \sum_i \left(-1+  \sum_{k=1}^i x_k \right) {\partial_{x_i}} \wedge{\partial_{\phi_i}} +\sum_{i>j}  x_i {\partial_{x_i}} \wedge{\partial_{\phi_j}} \,  . 
\end{equation}
As  the strong integrability and RG-flow analysis suggested, the second parameter $\zeta$ appears to be redundant. Indeed, we have explicitly checked up to $n=6$ that its  effect induced on metric and 3-form $H=\dd b$ can be removed  using new coordinates $\tilde{x}_i$ defined as
\begin{equation}
\tilde{x}_i = \frac{(k^2- \alpha^2) x_i}{[k + \alpha (2 \sum_{j< i} x_j -1)] [k + \alpha (2 \sum_{j \leq i} x_j -1)]}\, , \quad i=1, \dots, n 
\end{equation}
where $\alpha$ is given  by $\alpha^2 = k^2 -1 - \tilde{\eta}^2$ and leaving the angles $\phi_i$ untouched. Although we do not dispose of a general proof, we believe we can safely conjecture its existence for all $n$.  Given this diffeomorphism, we will henceforth drop any dependence on $\zeta$ setting $\zeta= 0$. Nevertheless, the geometry is still best described using the Poisson pencil  $\pi_\tau$ as in \eqref{eqn:PP} by fixing $\tau=1$.  

Even with a single parameter and in adapted coordinates, the forms of the deformed metric and $B$-field result in quite involved expressions and we will thus refrain from showing them here explicitly. Rather, we will argue in the next section that the resulting geometry is bi-Hermitian and we will provide canonical formulas for the geometric objects of interest.

\subsection{Generalised K\"ahler structure} \label{sec:gengeom}
We will now study the generalised K\"ahler structure of the above integrable models. We first present a rapid recap of the salient details of this geometry\cite{GualtieriThesis, Koerber}, before examining the  structure and pure spinors for the deformed \CP{n} keeping $n$ unspecified.   To progress further we give some explicit expressions for (generalised) K\"ahler potential for the cases of $n=1,2$.

\subsubsection{A pr\'ecis of Generalised K\"ahler Geometry } \label{sec:precis}
Demanding that a non-linear $\sigma$-model  admits an $N=(2,2)$ extended supersymmetry requires that the target space be bi-Hermitian \cite{Gates:1984nk,Sevrin:1996jr, Sevrin:2011mc}. That is, the  metric should be Hermitian with respect to two complex structures $J_\pm$ each of which is   covariantly constant with respect to the torsionful connections $ 0 = \nabla^{(\pm)} J_\pm = \left( \partial + \Gamma \pm H\right) J_\pm$ with  $H=\dd b$.  
As shown by   Gualtieri \cite{GualtieriThesis},  bi-Hermitian geometry is equivalent to  generalised K\"ahler geometry  in which the generalised metric can be decomposed in terms of two  commuting integrable generalised complex structures as  $\mathcal{E}  =\mathcal{J}_{1}\mathcal{J}_{2}$. 
The map between these two notions is 
\begin{equation}\label{eqn:gualmap}
	\mathcal{J}_{1,2} = \frac{1}{2} \begin{pmatrix} 1 &&0 \\ t^{-1} b && 1 \end{pmatrix} \begin{pmatrix}
		(J_+^t \pm J_-^t) &&- t ( \omega_+^{-1} \mp \omega^{-1}_-) \\
		t^{-1} (\omega_+ \mp \omega_-) && -(J_+ \pm J_-)
	\end{pmatrix}
	\begin{pmatrix}
		1 && 0 \\ - t^{-1} b && 1 
	\end{pmatrix} \, , 
\end{equation}with $\omega_\pm = J_{\pm} g$.

A useful alternative description is based on the construction of the pure spinors $\widehat{\Psi}_i$ associated to each complex structure. A generalised (not necessarily pure) spinor can be either viewed as chiral/anti-chiral Spin$(D,D)$ Majorana-Weyl spinor or as the formal sum of either even or odd degree differential forms \cite{GualtieriThesis}. When thought of as a  polyform, a spinor is naturally acted upon by a generalised vector $V = v+ \xi$  via the Clifford multiplication 
\begin{equation}
	V \cdot \widehat{\Psi} = \iota_v 	\widehat{\Psi} + \xi \wedge \widehat{\Psi} \,  .
\end{equation}
A spinor is then said to be \textit{pure} when its null space under the Clifford action is maximally isotropic. It can be shown that there is in fact a one-to-one correspondence between integrable generalised complex structures and non-degenerate ($\widehat{\Psi} \wedge \bar{\widehat{\Psi}} \neq 0$) complex pure spinors \cite{Koerber,Hitchin}. Integrability of $\mathcal{J}_i$ is recast as $\dd \widehat{\Psi}_i = \mathds{X}_i \cdot \widehat{\Psi}_i$, where $\mathds{X}_i$ are some generalised vectors. Finally, to a pure spinor we will associate an integer number $k$, called \textit{type},  corresponding to the lowest degree of the forms appearing in its expression.

Whichever point of view we wish to adopt, the entire geometrical data $(g,b,J_\pm)$ can {\em in principle} be encoded in  a single function $\mathcal{K}$ of superfields  called the  generalised K\"ahler potential \cite{Lindstrom:2007xv}.  However, explicit forms for such $\mathcal{K}$ are generally rather challenging to extract. 

To progress in this direction it is useful to first note that one can construct three Poisson structures\footnote{We choose to introduce an extra factor of $\pm 1/2$ with respect to the standard definitions so as to get rid of some numerical factors which will not affect the subsequent analysis.} \cite{Lindstrom:2005zr} $\pi_\pm =\pm 1/2 \left(\omega_+^{-1} \pm \omega_-^{-1} \right)$ and $\sigma=  g^{-1} [J_+, J_-] $. The types of supersymmetric multiplets required to furnish an $N=(2,2)$ action can be extracted from  these.    Namely, chiral superfields parameterise 
\begin{equation}\label{eq:kerJpminusJm}
 \ker(J_+ - J_- ) = \ker \pi_- \,  ,
\end{equation} 	
whereas twisted chirals are needed to parameterise 
\begin{equation}
\ker(J_+ + J_-) = \ker \pi_+ \,  .
\end{equation} 
 The remaining directions i.e. $\left(\ker([J_+, J_-]) \right)^\perp$ corresponding to the  symplectic leaves of $\sigma$  are to be parametrised by semi-chiral superfields \cite{Lindstrom:2005zr}.

A key challenge in establishing the generalised K\"ahler potential is  to find appropriate coordinates.   
  It is a trivial matter to check that $J^t_\pm \sigma J_\pm= - \sigma$, i.e. that $\sigma$ splits into $\sigma = \sigma^{(2,0)} + \overline{\sigma}^{(0,2)}$ with respect to either complex structures.  Invertibility, however, is not necessarily guaranteed. It is well known (see e.g. \cite{dufour2006poisson} for a comprehensive treatment) that each Poisson structure $\pi$ defines a foliation. Specifically, although $\pi$ might not be  globally  invertible, when restricted to one of its leaves $\Sigma$, the two-form $(\pi|_\Sigma)^{-1}$ is well-defined. 
   It has been first proven in \cite{Sevrin_1997} that for $\pi= \sigma$, the leaves have real dimension $4m$, for some $m \in \mathbb{N}$. In the  \CP{n} case, the integer $m$ is related to the complex dimension of the projective space via $m = [\frac{n}{2}]$.  
  
Suppose we now restrict to one leaf $\Sigma$, $\dim \Sigma = 4m$, where $\sigma^{-1}$ is well defined\footnote{Here and henceforth $\sigma^{-1}$ should be understood as $\sigma^{-1} \equiv (\sigma|_{\Sigma})^{-1}$.}. Because $\sigma$ is a Poisson structure $\dd \sigma^{-1} = 0$ has to hold. 
In general $\sigma^{-1}$ will inherit from $\sigma$ the decomposition $\sigma^{-1} = {\sigma^{-1}}^{(2,0)} + {\overline{\sigma}^{-1}}^{(0,2)}$  and the holomorphic coordinates we look for should be such that it is brought to the   canonical form
\begin{equation} \label{eq:sigmaDarboux}
  \sigma^{-1} = \sum_{i=1}^m\dd q^i \wedge \dd p_i + \mathrm{c.c.} =
   \sum_{i=1}^m \dd Q^i \wedge \dd P_i + \mathrm{c.c.}  
\end{equation} 
  Alternatively, $(q,p)$ and $(Q,P)$ can be thought of as the complex coordinates diagonalising, respectively, $J_+$ and $J_-$ restricted to $\Sigma$ (where they do not commute, so that they cannot be simultaneously diagonalised). In the language of supersymmetry, one can also look at $(q^i,\bar{q}^i,P_i, \bar{P}_i)$, for a fixed value of $i$, as part of a semi-chiral superfield\cite{Lindstrom:2007xv}.  
 The crucial point is that the transformation between $(p, \bar{p},Q, \bar{Q})$ and $(P,\bar{P},q, \bar{q})$   is canonical with a (real) generating function $\mathcal{K}(P,\bar{P},q, \bar{q})$ such that
 \begin{equation} \label{eqn:derK}
 p_i= \frac{\partial \mathcal{K}}{\partial q^i} \ , \quad  Q^i= \frac{\partial \mathcal{K}}{\partial P_i } \ . 
 \end{equation}
 It is this generating functional that becomes identified with the generalised K\"ahler potential. Thus extracting $\mathcal{K}$ can in general be hard and cumbersome: first one has to obtain $p_i$ and $Q^i$ and then integrate the above equations to determine $ \mathcal{K}$.

 The discussion above completely determines the K\"ahler potential when semi-chiral fields parametrise the whole geometry, as will turn out to be the case for the $\eta$-deformation of \CP{2m}.   As we shall see shortly, for \CP{2m+1} we will be required to augment the semi-chiral multiplets with a single chiral multiplet.   When chiral and/or twisted chiral multiplets are required, the algorithm for determining the K\"ahler potential is slightly more involved but has been detailed in the literature \cite{Lindstrom:2007xv,Zabzine_2009}.  In essence,   one simply  repeats the above construction on each symplectic leaf; however, the resulting expressions are somewhat more complicated  \cite{Lindstrom:2007xv}.  In the present paper, however, we will content ourselves with considering explicitly the K\"ahler potential for the case of \CP{1} and \CP{2}.
   
\subsubsection{\texorpdfstring{\CP{n}}{CPn}}  \label{sec:CPnresults}
We will start by studying the invariance property of \CP{n} as a coset manifold, taking $G=SU(n+1)$ and $H=S(U(n)\times U(1))$. In this case we obtain the branching rule
\begin{equation}\label{eqn:branchingSUn+1}
  \mathrm{adj}_G \rightarrow \mathrm{adj}_H + n_1 + \overline{n}_{-1}
\end{equation}
of $G$ irreps to $H$. Eventually our goal is to find forms which are invariant, i.e. singlets, under the holonomy group $H$. For $SU(n)$ we know that trivial representations only arise in the tensor product $n \times \overline{n}$. Thus, with the decomposition \eqref{eqn:branchingSUn+1}, we find that there are two $SU(n)$ invariant two forms: a symmetric one which is the restriction $\kappa$ of the Killing form on $G$ to the coset and an antisymmetric one which is $\omega$, the K\"ahler form. Both are invariant under $SU(n)$ and $U(1)$. Furthermore there are two $SU(n)$ invariant $n$ forms $\Omega$ and $\overline{\Omega}$ with $U(1)$-charges $n$ and $-n$, respectively. There also has to be an invariant $R_{\ca \cb}$ but the only invariant two-form we found is $\omega$. Thus, we conclude that $R_{\ca \cb}=\omega_{\ca \cb}$, in perfect agreement with the interpretation of $R_{\ca}{}^{\cb}$ as a complex structure on the algebra.

We can now observe that the generalised metric with flat indices given in eq.~\eqref{eq:genmet1} (now with $\zeta$ set to zero) admits the  decomposition $\mathcal{E}= \mathcal{J}_1 \mathcal{J}_2$  with  \begin{equation}\label{eq:flatcalJ}
{\mathcal{J}}_1{}_{\cA}{}^{\cB} = \begin{pmatrix}
    R^t & 0 \\
    0 & -R
  \end{pmatrix}
    \quad \text{and} \quad
{ \mathcal{J}}_2{}_{\cA}{}^{\cB} = \begin{pmatrix}
    \eta &&   (1+ \eta^2) t\kappa^{-1} R\\
  t^{-1} R \kappa && - \eta
  \end{pmatrix} \, , \end{equation}
   such that
\begin{equation}  
{  \mathcal{J}}_{i}^2 = - 1 \, , \quad [{  \mathcal{J}}_{1}, { \mathcal{J}}_{2}] = 0 \, . 
\end{equation}
Thus upon dressing these flat space quantities with the generalised frame fields \eqref{eq:genFFcoset} we see that the target space geometry is indeed   generalised K\"ahler with
\begin{equation} \label{eq:genCX}
{\cal J}_1  = \begin{pmatrix} J^t &t \eta (J^t \pi_{\mathrm{B}}+ \pi_\mathrm{B} J) \\ 0 & - J \end{pmatrix} \, ,  \quad {\cal J}_2  = \begin{pmatrix}- \eta \pi_{\tau}  \omega & - t (\omega^{-1} + \eta^2   \pi_{\tau} \omega \pi_{\tau} )  \\ t^{-1} \omega & \eta \omega \pi_{\tau}   \end{pmatrix}\, .
\end{equation}

It is now easy to show that, if we introduce the quantities $Q_\pm = 1 \pm b g^{-1}= 1 \mp \eta G \pi_\tau$, the objects appearing in the map \eqref{eqn:gualmap} are given by 
	\begin{equation}\label{eq:allpm}
		J_\pm = Q_\pm^{-1} J Q_\pm \:, \qquad \omega^{-1}_\pm = Q^t_\pm \omega^{-1} Q_\pm  \, , \qquad g = Q_\pm^{-1} G Q_\pm^{-t}\, .
	\end{equation}
The   three Poisson structures  $\pi_\pm$ and $   \sigma $  are extracted as  	\begin{equation}\label{eq:sigma}
		\pi_+ = ( 1+ \eta^2 \pi_{\tau} \omega \pi_{\tau} \omega) \omega^{-1} \, , \qquad \pi_- = \eta (J^t \pi_{\mathrm{B}}+  \pi_{\mathrm{B}}J )  \, , \qquad 	 \sigma = \omega_-^{-1}  J_+ - \omega_+^{-1} J_-  \, .
	\end{equation} 

	Let  us briefly study the superfields associated to these structures. For $\pi_+$, its kernel is isomorphic to the kernel of  $1+ (\eta \pi_{\tau} \omega)^2$ which, since $(\eta \pi_{\tau} \omega)^2$  is  positive definite, is trivial. Hence, no twisted chiral multiplets  are present.  The kernel for $\pi_-$ is better studied in Fubini-Study coordinates of the largest Bruhat cell. Here the complex structure is diagonal and  the expression \eqref{eq:sigma} for $\pi_-$ amounts to selecting the diagonal blocks of $\pi_{\mathrm{B}}$ which, in this patch, turn out the complex conjugates of one another. Each one of these blocks is a $n \times n$ dimensional matrix and, therefore, has vanishing determinant for odd $n$. In particular,  each block has a null space parametrised by one single  vector so that, upon linearly combining them, we have a total of two vectors generating the null space.   In the even case, it turns out that the determinant is non-vanishing, implying a trivial kernel. In summary, when $n$ is odd we have two vectors generating the kernel of $\pi_-$ and, thus, a single chiral superfield. We therefore end up with $(n-1)/2$  semi-chiral multiplets plus a single chiral multiplet  in the odd case and $n/2$ semi-chiral multiplets in the even case.  
	
	The same results can be obtained  employing pure spinors; we recall that the type $k_{1,2}$ of a pure spinor $\widehat{\Psi}_{1,2}$ is related to the kernel of the Poisson structures via the relation  
	\begin{equation} 
	\dim  \ker \pi_\mp = 2 k_{1,2} \, .
	\end{equation}
To compute these pure spinors it is efficient to first compute the corresponding flat space equivalents ${\Psi}_i$ corresponding to the complex structure  in eq. \eqref{eq:flatcalJ} and, using the spinorial representation of the generalised frame fields $S_E$, later build $\widehat{\Psi}_i = S_E{\Psi}_i$, where  $S_E = S_\pi S_e$ is the corresponding spinorial representatives of the decomposition in \eqref{eq:genFFcoset}. 
 
We can now check how the exterior derivative acts on these two spinors. Pulling the exterior derivate through the spinor action of the generalised frame field results in the $F$-twisted derivative 
 \begin{equation}
  \dd \widehat{\Psi}_{1,2} = S_E \dd_F {\Psi}_{1,2}\,, \quad \text{with} \quad 
  \dd_F  = - \frac14 e^{-d} \Omega_{\cA\cB\cC} \Gamma^{\cA} \Gamma^{\cB\cC} \, 
\end{equation}
with the generalised dilaton $d=\Phi - 1/4 \log \det g$. Note that because we are considering a symmetric space there is no $F_{ABC}$  contribution. Using the definition \eqref{eqn:OmegaalphaB} of the generalised connection we eventually obtain
\begin{equation} \label{eq:spinorintegrability}
  \dd \widehat{\Psi}_1 = 0 \qquad \text{and} \qquad \dd \widehat{\Psi}_2 = \mathds{X} \cdot\widehat{\Psi}_2
  \qquad \text{with} \qquad \mathds{X} = e^{-d} \left( e^0 + \pi^{0\ca} v_{\ca} \right)\,,
\end{equation}
where the index $0$ refers to the component of $e^\alpha$ and the bi-vector along the holomorphic $U(1)$ in the branching \eqref{eqn:branchingSUn+1}. This proves that we have a generalised K\"ahler manifold (eq. \eqref{eq:spinorintegrability} is equivalent to the integrability of the generalised complex structures, cf. Section \ref{sec:precis}) 
and at the same time that the three Poisson structures we presented above are indeed Poisson structures.

Finding the generalised K\"ahler potential $\mathcal{K}$ for a deformation of \CP{n}  is complicated, at least for  generic $n$. We thus defer to the next section the explicit computation in the $n=1,2$ cases, whilst we devote the present paragraph to some more general reasoning.  To find the $(p,q)$ and $(P,Q)$ coordinates explicitly we exploit the fact that there are $n$ Killing vectors which leave $\sigma^{-1}$ invariant (considering for simplicity here the case relevant to \CP{even} for which $\sigma$ is invertible).  The  coordinates $(x_m, \phi_m)$ introduced in eq.~\eqref{eqn:xphi} are adapted to this such that the Killing vectors are  simply given by the  $\partial_{\phi_m}$.  
We can select the holomorphic (with respect to $J_\pm$) part of $\sigma^{-1}$ by acting with a projector
\begin{equation}
  \sigma^{-1}_\pm = \frac1{2 i} ( i + J_\pm ) \sigma^{-1} \, , 
\end{equation}
such that 
\begin{equation}
  \dd q \wedge \dd p + \dd Q \wedge \dd P = \sigma_+^{-1} + \sigma_-^{-1} \,.
\end{equation}
Because both $\sigma^{-1}_\pm$ are invariant under the action of the Killing vectors $\partial_{\phi^m}$
\begin{equation}
  L_{\partial_{\phi^m}} \sigma^{-1}_\pm = 0 = \dd ( \iota_{\partial_{\phi^m}} \sigma^{-1}_\pm ) \, , 
\end{equation}
we obtain the momentum maps
\begin{equation}\label{eqn:momentummap}
  \dd \mu_m^\pm = \iota_{\partial_{\phi^m}} \sigma^{-1}_\pm  \, , 
\end{equation}
which, together with the one-forms $\dd \phi^m$ dual to the isometries, form a basis  of one-forms. A symplectic form $\sigma^{-1} $ which satisfies \eqref{eqn:momentummap} has to have the form  
\begin{equation}
  \sigma_{\pm }^{-1} = \frac12 ( a + a b a )_{mn} \dd \phi^m \wedge \dd \phi^n + ( 1 + a b )_m{}^n \dd \phi^m \wedge \dd \mu^\pm_n + \frac12 b^{mn} \dd \mu^\pm_m \wedge \dd \mu^\pm_n\, ,
\end{equation}
where
\begin{equation}
  a_{mn} = \iota_{\partial_{\phi^m}} \dd \mu_n
  \quad \text{with} \quad \dd a_{mn} = 0\,.
\end{equation}
Furthermore, $\sigma^{-1}$ has to be closed. This implies that the only free parameter $b^{mn}$ has to be  constant like $a_{mn}$. To fix $b^{mn}$, we just have to match the left and right hand side. As result, we find that
\begin{equation}
  \sigma^{-1}_+ = \dd q^m \wedge \dd p_m \quad \text{and} \quad
  \sigma^{-1}_- = \dd Q^m \wedge \dd P_m \,,
\end{equation}
where $\dd q^m$ and $\dd p_m$ are linear combinations (with constrained coefficients) of $\dd\phi^m$ and $\dd \mu_m^+$. The same holds for $\dd Q^m$ and $\dd P_m$ but with respect now to the linear combination built from $\dd \phi^m$ and $\dd \mu_m^-$.     So the procedure is simple in principle: first  integrate the moment map to find the $\mu_m$ and take appropriate linear combinations $\mu$ and $\phi$ to define the canonical coordinates. Then find the generating function $\cal K$ by integrating   the canonical transformation of eq.~\eqref{eqn:derK}.

\subsubsection{Pure spinors}

Pure spinors for $\eta$-deformed \CP{n} can be studied without fixing a specific (complex) dimension. We will follow the standard procedure, namely we will first compute a basis $V_{1,2}^j$ for the $+i$-eigenspace of each complex structure and then we will impose that the same basis annihilates the associated pure spinor. 

Let us start from $\mathcal{J}_1$. It is most easily analysed in Fubini-Study coordinates: $J$ is diagonal and $n$ $+i$-eigenvectors for $\mathcal{J}_1$ are immediately found to be $V^j_1 = \partial_{z_j}$, $j=1, \dots, n$. On the other hand, $\pi_-$ in these coordinates reads
\begin{equation}
 \pi_- = 2 \eta \sum_{j>i} (z_i z_j \partial_{z_i} \wedge \partial_{z_j} + \mathrm{c.c.}) \,  ,
\end{equation}
making it easy to see that the remaining eigenvectors are
\begin{equation} \label{eq:gensecondset}
 V_1^{n+j} = \dd \bar{z}_j + i \eta t \, \bar{z}_j \left(\sum_{i > j} \bar{z}_i \partial_{\bar{z}_i} -\sum_{i < j} \bar{z}_i \partial_{\bar{z}_i} \right) \, , \qquad j=1, \dots, n \,  . 
\end{equation}
As proved by Gualtieri \cite{GualtieriThesis}, the general form of a non-degenerate complex pure spinor is $\widehat{\Psi} = \Xi \wedge e^{\rho}$, where $\rho$ is a complex two-form, $\Xi$ a decomposable $k$-form and $k$ the type of the spinor.  For odd $n$, we proved that $\dim \ker \pi_- = 2$ and the spinor will be of type 1 (that is, $\Xi$ will be a one-form). On the contrary, for even $n$ the Poisson structure $\pi_-$ has trivial kernel: the spinor will have type 0 and we can consistently set $\Xi=1$ since the spinor is defined up to an overall function. 

Now, the requirement $V_1^j \cdot \widehat{\Psi}_1 = 0$ for $j=1, \dots, n$ implies that the spinor is made up of anti-holomorphic forms only. The constraints arising from $V_1^{n+j} \cdot \widehat{\Psi}_1 = 0$ with $j=1, \dots, n$ are equivalent to 
\begin{align}
		0 = \xi_1^{n+j}  + \iota_{v_1^{n+j}} \rho  \qquad &\text{for even } n \, ,\\
		\label{eq:secondpsi1}
		0 = \iota_{v_1^{n+j}} \Xi - \Xi \wedge \iota_{v_1^{n+j}} \rho + \xi_1^{n+j} \wedge \Xi \qquad &\text{for odd } n \, , 
\end{align}
where $v_1^{n+j}$ and $\xi_1^{n+j}$ are, respectively, the vector and form part of the generalised vectors \eqref{eq:gensecondset}. Observe that \eqref{eq:secondpsi1} can be in fact split into two separate equations, corresponding to degree zero and two. In this sense, the degree zero requirement is the same as saying that the interior product of $\Xi$ with $v_1^{n+j}$ vanishes for all $j=1, \dots, n$. 
As one can explicitly check, all of the equations are satisfied with
\begin{equation}
 \rho = \frac{i}{\eta t} \sum_{k>i} (-1)^{i+k} \frac{\dd \bar{z}_i \wedge \dd \bar{z}_k}{\bar{z}_i \bar{z}_k}	\qquad \text{and} \qquad \Xi = \begin{cases} 	1 & \text{even } n \\
 	i \eta t \sum_{k} (-1)^{k+1} \frac{\dd \bar{z}_k}{\bar{z}_k} & \text{odd } n  
 	\end{cases}
  \,  .
\end{equation}
With this normalisation, we remark that for vanishing $\eta$ the pure spinor is well defined and coincides (after an appropriate rescaling) with the decomposable anti-holomorphic form $\overline{\Omega} = \dd \bar{z}_1 \wedge \dots \wedge \dd \bar{z}_n$. 

As for $\mathcal{J}_2$ it is sufficient to notice that its explicit form \eqref{eq:genCX} implies that each and every generalised eigenvector $V_2^j$ with $+i$ eigenvalue  will be given by
\begin{equation}
	V_2^j = i t (\omega^{-1}+ i \eta \pi_\tau) \xi_2^j + \xi_2^j \, \qquad j=1, \dots, 2n  ,
\end{equation}
being $\xi_2^j$ a set of $2n$ independent one-forms. The second pure spinor then results in
\begin{equation}
	\widehat{\Psi}_2 = \exp \left[- i t^{-1} (\omega^{-1}+ i \eta \pi_\tau)^{-1}\right] \,  .
\end{equation}
In particular, notice that the $\eta \rightarrow 0$ limit correctly yields the exponential of the K\"ahler form, as it should for a K\"ahler manifold. 

Finally, notice that, for every value of $n$, $\dd \rho = \dd \Xi =0$; also, $\dd (\omega^{-1}+ i \eta \pi_\tau)^{-1} =0$ follows from the compatibility of $\pi_{\mathrm{B}}$ and $\omega^{-1}$. Thus, $\dd \widehat{\Psi}_{1,2} = 0$. Actually, this is a consequence of our choice of normalisation for the spinors; for instance, we have set the zero-form component of $\widehat{\Psi}_2$ to one. Instead, we could impose a different normalisation using the Mukai pairing $|| \widehat{\Psi}_{i}||^2 = \widehat{\Psi}_i \wedge \sigma (\bar{\widehat{\Psi}}_i)|_{\mathrm{top}}$, where $\sigma$ reverses the order of the indices in the polyform and $|_{\mathrm{top}}$ stands for restriction to the top-dimensional form. Should we scale the pure spinors such that they have equal normalisation, then they would no longer be closed. The geometry is hence not generalised Calabi-Yau\footnote{A generalised K\"ahler geometry is generalised Calabi-Yau when the pure spinors associated to the generalised complex structures are nowhere-vanishing, closed when choosing their relative norm with respect to the Mukai-pairing to be a constant \cite{GualtieriThesis, Koerber}.}. 

\subsubsection{Generalised K\"ahler potential}

\paragraph{\texorpdfstring{\CP{1}}{CP1}}

It is a well-known fact that every two-dimensional complex manifold is K\"ahler; as such, the  deformed \CP{1} geometry is completely determined by the standard (i.e. non generalised)  K\"ahler potential. In fact, one can further notice that, given the dimensionality, the $B$-field is always pure gauge and thus negligible.
 As for the patch, we put ourselves in the largest Bruhat cell where the homogeneous coordinate $Z_0 \neq 0$ and introduce the holomorphic coordinate $z \equiv Z_1/Z_0$. The K\"ahler potential is 
\begin{equation}
  K = -\frac{1}{2 \eta} \mathrm{Im} \, \mathrm{Li}_2 \left(\frac{\eta - i}{\eta + i} |z|^2  \right)  \, , 
\end{equation}
where we  notice  that the $\eta \rightarrow 0$ limit is non-singular and yields $ K_{\mathrm{FS}}$,  i.e. the  undeformed Fubini-Study K\"ahler potential for \CP{1}, $K_{\mathrm{FS}}= \log (1+|z|^2)$. 

As \CP{1} is K\"ahler, $J_+ = J_-$, and $\pi_-$ vanishes. In turn there is a single set of complex coordinates diagonalising $J_\pm$ expressed by 
\begin{eqnarray}
q = - 2 \mu \log (z)= \mu \left( \log\left( \frac{1-x}{x} \right)  - 2 i\phi \right) = \mu \left( \log\left(  \sin(\beta + \chi) \csc(\beta- \chi) \right)  - 2 i\phi \right)  \, ,
\end{eqnarray}
and its conjugate, where $(x, \phi)$  were introduced in \eqref{eqn:xphi},  and\footnote{Strictly speaking, for \CP{1} the precise form for  $\mu$ is undetermined. We nevertheless choose it so as to match the  higher dimensional cases, see next section.}
\begin{equation} \label{eq:chi1}
x = \frac{1}{2 \eta}(\eta - \tan \chi) \, , \qquad \mu = \frac{i -1}{8 \sqrt {\eta \,  t}} \, , \qquad \eta = \tan \beta \,  .
\end{equation}

\paragraph{\texorpdfstring{\CP{2}}{CP2}} 
\CP{2} is the first case where we can study a non-trivial generalised K\"ahler potential and give a rather nice explicit presentation thereof.

 A  first step in computing it  is to find the holomorphic coordinates of $J_\pm$, that is, to identify  $p(z,\bar{z}), q(z,\bar{z}),P(z,\bar{z}), Q(z,\bar{z})$, such that
\begin{equation}\label{eq:JpJms}\begin{aligned}
J_+  =&~i  \dd p \otimes \partial_{p} -i \dd \bar{p} \otimes \partial_{\bar{p}} +i  \dd q \otimes \partial_{q} -i  \dd \bar{q} \otimes \partial_{\bar{q}} \, ,  \\ 
J_-  =&~i   \dd P \otimes \partial_{P} -i  \dd \bar{P} \otimes \partial_{\bar{P}} +i  \dd Q \otimes \partial_{Q} -i \dd \bar{Q} \otimes \partial_{\bar{Q}} \, ,  \\
\sigma^{-1} =&~\dd p \wedge \dd q +  \dd \bar{p} \wedge \dd \bar{q} = \dd P\wedge \dd Q + \dd \bar{P} \wedge \dd \bar{Q} \, .  
\end{aligned}\end{equation}
Using the   symplectic moment map  associated to  $U(1)$ actions as described previously one finds 
for $p,q$ (with $\bar{p},\bar{q}$ given by standard complex conjugation)
\begin{equation} \label{eq:qpdef}
\begin{aligned} 
q=&
\mu  \left(\log \left(-e^{-i \chi _2} \sin \left(\beta +\chi _1-\chi _2\right) \csc \left(\beta -\chi
_1\right)\right)- 2 i \phi _1 \right)\, , \\
p =& \mu \left(  \log \left(-i e^{-i \chi _1} \sec (\beta ) \csc \left(\chi _2\right) \sin \left(\beta +\chi
_1-\chi _2\right)\right)-2 i \phi _2 \right) \,  ,
\end{aligned}
\end{equation}
where the angles $\chi_{1,2}$ are a generalisation of the one previously  introduced 
\begin{equation}\label{eq:chiangles}
x_1= \frac{1}{2} - \frac{1}{2\eta} \tan(\chi_1)\, , \quad  x_2= \frac{1}{2\eta}\sec(\chi_1) \sec  ( \chi_1- \chi_2 )\sin(\chi_2) \, ,
\end{equation}
and $\mu$ and $\beta$ follow the definition in \eqref{eq:chi1}. In particular,  $\mu $ is  a  coefficient needed to  ensure that $\sigma^{-1}$ has the correct form \eqref{eq:JpJms}. The relations \eqref{eq:qpdef} in Fubini-Study coordinates are
\begin{align}
	q &= \mu \left[- 2 \log \left(\frac{\sqrt{z_1}}{\sqrt{\bar{z}_{1}}}\right) + \log \left(-\frac{\eta(1-|z_1|^2+|z_2|^2) + i (1+|z|^2)}{|z_1|^2 (\eta (1-|z|^2)+ i (1+|z|^2))}\right)  \right]\,  , \\
	p &= \mu \left[- 2 \log \left(\frac{\sqrt{z_2}}{\sqrt{\bar{z}_{2}}}\right) + \log \left(- \frac{\eta(1-|z_1|^2+|z_2|^2)+ i (1+|z|^2)}{|z_2|^2 (1+|z|^2)}\right)\right] \,  ,
\end{align} 
where we recall $|z|^2 \equiv |z_1|^2 + |z_2|^2$. 
Instead, one can use the angles $\chi_{1,2}$ to show that a simple relation between $p,q$ and $P,Q$ exists, namely
\begin{equation} \label{eq:qp}
p +P=   -  2 i \mu \chi_1 \,, \qquad q+ Q =   -  2 i \mu \chi_2\, .
\end{equation}
Letting  the generating function
\begin{equation}
\mathcal{K}(P,\bar{P},q, \bar{q}) = -(P q  + \bar{P} \bar{q}) + \mathcal{K}_1(P,\bar{P},q, \bar{q})  \, , 
\end{equation}
we require, in accordance with \eqref{eqn:derK} and \eqref{eq:qp}, that 
\begin{equation}
 \partial_{q}\mathcal{K}_1=-  2 i    \mu \chi_1  \ , \quad  \partial_{P} \mathcal{K}_1 =  -2 i  \mu \chi_2 \,. 
\end{equation} 
A closed form for the potential in terms of the angles $\chi_i$ can be given in terms of  the parametric integral
\begin{equation}
{\cal I}_\alpha (y) = \int y \cot \left(\frac{y +\alpha}{2} \right)  \dd y = 2 \left(y \log \left(1-e^{i (\alpha +y)}\right)-i \text{Li}_2\left(e^{i (y+\alpha
	)}\right)\right)-\frac{i y^2}{2} \, , 
\end{equation} 
such that 
\begin{equation}
\mathcal{K}_1(\chi_1, \chi_2) =  \frac{1}{32 t \eta} \left( {\cal I}_{-2\beta}  (2 \chi_1)  - {\cal I}_{2\beta}( 2 \chi_1 - 2\chi_2) - {\cal I}_0 (2 \chi_2)\right)\, . 
\end{equation} 
To complete the specification of the potential one needs to express the $\chi_i$ in terms of $(P,\bar{P},q, \bar{q})$ which can be done implicitly via the relations
\begin{equation}
\begin{aligned}
e^{|p|/ |\mu| } =& \sec ( \beta ) \csc(\chi_2  )\sin(\beta + \chi_1 - \chi_2)    \, ,  \\ 
e^{|q|/ |\mu| } =& \csc(\beta - \chi_1) \sin(\beta + \chi_1 - \chi_2)   \, . 
\end{aligned} 
\end{equation} 

\section{Conclusions and outlook}
In this work, using the tools of Poisson-Lie non-linear $\sigma$-models on generalised coset spaces $G/H$ \cite{Demulder:2019vvh}, we have described a specific but particularly striking class of examples in which $G/H$ were Poisson Hermitian spaces. Upon constructing an integrable Yang-Baxter deformation of these, we showed that their target space is described by generalised K\"{a}hler geometry.   
We discussed in detail \CP{n} as a prototypical example and, for the case of  \CP{2}, gave an explicit formulation of the corresponding generalised K\"{a}hler potential.  We filled a gap in the literature by showing that a previously conjectured two parameter deformation for \CP{n} is indeed integrable but coinciding with the already well-known Yang-Baxter deformation of coset spaces. 

A background motivation for this work was to investigate integrable deformations of  $AdS_4\times   \text{\CP{3}}$ with the aspiration of identifying quantum group deformations in the ABJM model. At first sight the corresponding geometry is rather unattractive but in this work we have elucidated many of the key features.  A complete analysis would of course require furnishing the geometry with appropriate RR fields and investigating the fermionic sector.     Whilst one might ``boot-strap'' an RR sector, an approach done for the case of Poisson-Lie models on groups in \cite{Demulder:2018lmj}, ultimately it would be desirable to extend the considerations  to super-cosets \cite{Hoare:2014pna, Hoare:2011wr, Hoare:2015gda, Hoare:2018ebg, Hoare:2018ngg}.  

In this letter we only considered coset spaces for which the gauge group is coisotropic, as these naturally solve the invariance constraint, leaving the construction of other holographically relevant coset spaces open. Moreover, the explicit examples taken into account here were based on quotients in which the Drinfel'd double was $\mathfrak{d} = \mathfrak{g}^\mathbb{C}$. The incorporation of $\lambda$ models requires the more general case of $\mathfrak{d}= \mathfrak{g}+\mathfrak{g}$; our general tool kit \cite{Demulder:2019vvh} accommodates this scenario and so it would be interesting to explore if there can be some underlying generalised K\"ahler structures in the $\lambda$-deformations of $G/H$-WZW models.   

A fruitful further line of investigation would be to consider the presence of D-branes  in these geometries and understand how the elegant characterisation in generalised K\"ahler geometry can relate to   integrability preserving boundary conditions as in e.g. \cite{Driezen:2018glg}. Beyond such direct follow up tasks, this work suggests a number of interesting broader questions, and we list a few here with the hope of returning to them later:
\begin{itemize}
\item How generic are the constructions of Poisson-Lie models on groups and coset spaces; how much of the landscape of integrable  $\sigma$-models do they capture?
 \item We saw some interplay between integrability and extended supersymmetry. However one need not expect all $N=(2,2)$ models to be integrable, but which are?  How is this reflected at the level of superspace? 
\end{itemize}

\section*{Acknowledgements}

We gratefully acknowledge N. Ciccoli for useful correspondence. DCT is supported by a Royal Society University Research Fellowship {\em Generalised Dualities in String Theory and Holography} URF 150185 and in part by STFC grant ST/P00055X/1 and in part by the ``FWO-Vlaanderen'' through the project G006119N and by the Vrije Universiteit Brussel through the Strategic Research Program ``High-Energy Physics''.   SD acknowledges Max-Planck-Society for support. GP is supported by a Royal Society Enhancement Award RGF/EA/180176. The work of FH is supported in part by DOE grant DE-FG02-13ER42020 and the Cynthia and George Mitchell Foundation.

\appendix
 \section{Coisotropic subgroups and Poisson pencils} \label{app:coisotropy} 
In this section we provide a short introduction to the notions of coisotropic subgroup and Poisson pencil, as they are extensively used in the paper; we will not aim at a comprehensive treatment and instead refer the interested reader to standard textbooks on the subject, e.g. \cite{Chari}.

We take $\mathfrak{g} = \mathrm{Lie}(G)$ to be a semi-simple Lie algebra. As such, $\mathfrak{g}$ is naturally endowed with a non-degenerate pairing $\kappa$, the Killing form. Non-degeneracy implies that $\kappa$ induces an isomorphism $\phi : \mathfrak{g} \rightarrow \mathfrak{g}^*$ explicitly given by 
\begin{equation}
\phi(x)  = \kappa(x, \cdot) \in \mathfrak{g}^* \quad \forall x \in \mathfrak{g} .
\end{equation}
We would like to promote $\mathfrak{g}$ to a bialgebra which, by virtue of the Whitehead's lemma, is completely specified by a skew-symmetric Yang-Baxter matrix $r \in \mathfrak{g} \otimes\mathfrak{g}$ obeying the modified classical Yang-Baxter equation. The matrix induces a Lie bracket $[\cdot, \cdot]_*$ on the dual space $\mathfrak{g}^*$
\begin{equation}
[\xi, \eta]_* = \mathrm{ad}^*_{r \xi} \eta - \mathrm{ad}^*_{r \eta} \xi \,  ,
\end{equation}
where $\mathrm{ad}^*$ is the coadjoint action and $\xi, \eta \in \mathfrak{g}^*$. 
Alternatively, a Yang-Baxter matrix can be seen as an endomorphism $R: \mathfrak{g} \rightarrow \mathfrak{g}$ obeying the modified classical Yang-Baxter equation 
\begin{equation} \label{eq:mCYBEApp}
[Rx, Ry] = R([Rx,y]+[x,Ry]) -c^2 [x,y]  \qquad \forall x,y \in \mathfrak{g}\, ,
\end{equation}
where $c^2$ is a parameter which can be taken to be either $\pm 1$ or $0$. Here, we will fix $c^2=-1$. Once a basis $\{T_\ga\}$ for $\mathfrak{g}$ is specified, the two matrices are  related by the action of the Killing form, $r^{\ga \gb} = \kappa^{\ga \gc} R_\gc{}^\gb$. 
Defining a new operation $[\cdot, \cdot]_R : \mathfrak{g} \otimes \mathfrak{g} \rightarrow \mathfrak{g}$ given by
\begin{equation}
[x,y]_R = [Rx, y]+ [x, Ry] \, , 
\end{equation}
one can show that, if $R$ obeys the mCYBE \eqref{eq:mCYBEApp}, $[\cdot, \cdot]_R$ is a Lie bracket. Hence,  $\mathfrak{g}$ can be equipped with two different brackets, giving rise to two sets of structure constants
\begin{equation}
[T_\ga, T_\gb] = f_{\ga \gb}{}^\gc T_\gc \, , \qquad [T_\ga, T_\gb ]_R= \tilde{f}_{\ga \gb}{}^{\gc} T_\gc = -2 R_{[\ga}{}^\gd f_{ \gb]\gd}{}^\gc T_{\gc} \, .
\end{equation}
The brackets $[\cdot, \cdot]_*$ and $[\cdot, \cdot]_R$ are actually related. 
Indicating with $\langle \cdot, \cdot \rangle$ the canonical pairing between $\mathfrak{g}$ and $\mathfrak{g}^*$ we have, for $x \in \mathfrak{g}$
\begin{equation} \label{eq:bracketrelations}\begin{gathered}
\langle [\phi(z), \phi(y)]_*, x \rangle = \kappa([z,y]_R, x) \,  .
\end{gathered}\end{equation}

A subalgebra $\mathfrak{h}$ of a Lie bialgebra $\mathfrak{g}$ is called \textit{coisotropic} if its annihilator $\mathfrak{h}^\bot$, i.e. the space of functionals $\xi \in \mathfrak{g}^*$ such that$\langle \xi, x \rangle=0$ $\forall x \in \mathfrak{h}$, is a Lie subalgebra in $\mathfrak{g}^*$ \cite{LuThesis}.

Take $M= \exp(\mathfrak{m})$ to be the coset $M=G/H$ and further require it to be a symmetric space, so that $\mathfrak{g} = \mathfrak{m} \oplus \mathfrak{h}$  and  $\kappa(\mathfrak{m}, \mathfrak{h}) =0$, where $\mathfrak{h}$ and $\mathfrak{m}$ are, respectively, the $+1$ and $-1$ eigenspace of the $\mathbb{Z}_2$ involution. Defining $\mathfrak{m}^* = \phi(\mathfrak{m})$ and $\mathfrak{h}^* = \phi(\mathfrak{h})$ (so that $\mathfrak{g}^* = \mathfrak{m}^* \oplus \mathfrak{h}^*$), orthogonality implies $0=\kappa(\mathfrak{m}, \mathfrak{h}) = \langle \mathfrak{m}^*, \mathfrak{h} \rangle$. There can be no $\xi \in \mathfrak{h}^*$ obeying  $\langle \xi, \mathfrak{h} \rangle = 0$, or otherwise the restriction of $\kappa$ to  $\mathfrak{h}$ would be degenerate,  hence $\mathfrak{h}^\perp = \mathfrak{m}^*$. Without further constraints, $\mathfrak{m}^*$ is not a subalgebra of $\mathfrak{g}^*$, as needed for $\mathfrak{h}$ to be coisotropic. Requiring $[\mathfrak{m}^*, \mathfrak{m}^*]_* \subset \mathfrak{m}^*$ is equivalent to imposing $[\mathfrak{m}, \mathfrak{m}]_R|_{\mathfrak{h}} = 0$, via \eqref{eq:bracketrelations}. We obtain the coisotropy condition
\begin{equation}
([Rx, y]+[x,Ry])|_{\mathfrak{h}} = 0 \qquad \forall x,y \in \mathfrak{m} \,  .
\end{equation}

If $\mathfrak{h}$ is coisotropic and the coset $G/H$ is a symmetric space, $H^* = \exp(\mathfrak{h}^*)$ is a subgroup of $G^* = \exp(\mathfrak{g}^*)$ and the coset $M^* =G^*/H^*$ is a symmetric space, provided we endow $\mathfrak{g}^*$ with the Lie bracket $[\cdot, \cdot]_*$ induced by the Drinfel'd-Jimbo R-matrix\footnote{In general, the Drinfel'd-Jimbo procedure is not the unique possibility for constructing an R-matrix. However, it is most useful when building Poisson bi-vectors out of Yang-Baxter matrices, as in our case.}. This is most easily seen using the dual bracket $[\cdot, \cdot]_R$.  More precisely, grading $\mathfrak{g}$ into $\pm 1$-eigenspaces, it follows from the definition of the Cartan-Chevalley basis that, for a fixed root $\lambda$, the ladder operators $X_\lambda$ and $X_{-\lambda}$ belong to the same subspace, while the Cartan subalgebra  belongs to the $+1$ eigenspace. The Drinfel'd-Jimbo construction then implies that the Yang-Baxter matrix has no mixed components, $R(\mathfrak{h})|_{\mathfrak{m}} = 0$ and $R(\mathfrak{m})|_{\mathfrak{h}} = 0$. This fact, together with coisotropy and symmetric space decomposition, yields
\begin{equation} \label{eq:symmspace2}
[\mathfrak{m}, \mathfrak{m}]_R = 0 \, , \qquad [\mathfrak{m}, \mathfrak{h}]_R \subset \mathfrak{m} \, , \qquad [\mathfrak{h}, \mathfrak{h}]_R \subset \mathfrak{h} \,  .
\end{equation}
Lifting these conditions to the dual algebra $\mathfrak{g}^*$ we get 
$[\mathfrak{h}^*, \mathfrak{h}^*]_* \subset \mathfrak{h}^*$, $[\mathfrak{h}^*, \mathfrak{m}^*]_* \subset \mathfrak{m}^*$ and  $[\mathfrak{m}^*, \mathfrak{m}^*]_*  = 0$, the defining relations for a (particular type of) symmetric space $M^*$. For instance, it can be checked for $G/H = SU(2)/U(1)$: assuming $\mathfrak{m} = \mathrm{Span}(\sigma_1, \sigma_2)$, where $\sigma_i$ are the Pauli matrices, the Drinfel'd-Jimbo R-matrix acts as $R(\sigma_1) = \sigma_2$, $R(\sigma_2)= - \sigma_1$ and $R(\sigma_3)=0$; the relations \eqref{eq:symmspace2} follow. 

\color{black}  Verifying that a given $H$ is coisotropic, or better said coisotropically embedded, in $G$ is no trivial task. One possibility  is to exploit the method introduced in \cite{Zambon}; however, we shall pursue a different approach which better reflects the (Poisson-Lie) group theoretic origin of our system.  

A subgroup $H$ of a Poisson-Lie group $G$ is called a Poisson-Lie subgroup if the annihilators $\mathfrak{h}^\bot$ are an ideal in $\mathfrak{g}^*$\cite{LuThesis}. Since any ideal is a subalgebra, any Poisson-Lie subgroup is automatically a coisotropic subgroup (but the converse is in general false). Moreover it will be useful for us to notice that, whenever $H$ is a Poisson-Lie subgroup, the reduction of the Poisson structure $\pi$ defined on $G$ to its  restricted counterpart on $G/H$ is unique and gives rise to the so-called Poisson-Bruhat structure $\pi_{\mathrm{B}}$ \cite{Chari}. 

Another relevant fact  is  that every coadjoint orbit $\mathcal{O}$ of a Poisson-Lie group arises from the quotient of the latter with a Poisson-Lie subgroup\cite{LuWeinstein}.  Thus  coadjoint orbits of (compact semi-simple) Poisson-Lie groups can serve useful examples of target spaces in the theories we are considering.

Such coadjoint orbits  can  be equipped with a closed, non-degenerate, symplectic 2-form $\omega$ usually known as the Kostant-Kirillov-Souriau (KKS) form, where non-degeneracy implies that each orbit is an even-dimensional symplectic manifold. It can be further shown \cite{Borel, Besse} that this symplectic structure is actually a K\"ahler form and thus every $\mathcal{O}$ is a K\"ahler homogeneous space. \\
Now, in the light of the above, on each orbit we evidently have at our disposal two different Poisson structures, namely $\pi_\mathrm{B}$ and $\omega^{-1}$, since $\mathrm{d} \omega = 0$ implies $[\omega^{-1}, \omega^{-1}]_s =0$ (where $[\cdot, \cdot]_s$ is the Schouten bracket).  Given two Poisson structures $\pi_1$ and $\pi_2$ one could try and  combine them so as to create a third one, $\pi(\tau) = \pi_1 + \tau \pi_2$, where $\tau \in \mathbb{R}$. In general, the Schouten bracket $[\pi(\tau), \pi(\tau)]_s$ does not vanish but whenever $[\pi_1, \pi_2]_s= 0$ (and in this case $\pi_1$ and $\pi_2$ are said to be \textit{compatible}) it obviously does. If such a condition is met then $\pi(\tau)$ is referred to as a \textit{Poisson pencil}.  On the coadjoint orbit $\mathcal{O}$, checking whether $[\pi_{\mathrm{B}}, \omega^{-1}]_s = 0$ is not straightforward and might require an explicit calculation. However, as shown in \cite{khoroshkin1993}, on any (compact) Hermitian symmetric space the two Poisson structures $\pi_{\mathrm{B}}$ and $\omega^{-1}$ are compatible and hence give rise to a pencil.

\subsubsection*{Poisson pencils on Hermitian symmetric spaces}
Thus we are particularly interested in finding coadjoint orbits which are also Hermitian symmetric spaces so that they will be naturally endowed with a Poisson pencil. The first step is to observe, following Kirillov \cite{Kirillov}, that any coadjoint orbit is a \textit{generalised flag manifold}\footnote{Equivalently, Lu and Weinstein \cite{LuWeinstein} have shown that coadjoint orbits can be obtained as the quotient $\mathcal{O}= G^{\mathbb{R}}/P$, being $G^{\mathbb{R}}$ the complexified (algebraic) group over the reals and $P$ a parabolic subgroup thereof.}, a particular form of algebraic varieties. On the other hand, compact Hermitian symmetric spaces are well-known to be classified. One class of manifolds lying at the intersection of these two are the complex Grassmanians, given by the quotient $SU(n+m)/S(U(n) \times U(m))$ which includes as   the $m=1$ case  the complex projective spaces $\mathbb{C}\mathrm{P}^n$.

Being complex manifolds, they come with an integrable complex structure $J$. For a given choice of the parameter $c$, namely $c^2=-1$, the mCYBE coincides with the condition of vanishing Nijenhuis tensor: this observation provides an intuitive hint to the fact that the $R$-matrix on $\mathbb{C}\mathrm{P}^n$, thought of as the coset space above, could be interpreted as a complex structure. This statement has been made precise by Koszul \cite{SB_1954-1956__3__69_0} who has proved that, in order for $R$ to be a complex structure (in flat indices) on a reductive coset $\mathfrak{m}$ (where $M=G/H$), it is required to satisfy, besides the mCYBE (now interpreted as an integrability condition), the constraints
\begin{equation}
J|_\mathfrak{m}^2 = -1 \:, \qquad J \, \mathfrak{h} \subset \mathfrak{h} \: , \qquad [\mathfrak{h}, J \mathfrak{m}] = J[\mathfrak{h}, \mathfrak{m}] \: . 
\end{equation}
In addition, analogously to the Darboux theorem for standard complex structures, the $R$-matrix can be diagonalized\cite{Liu}: if we consider the two subalgebras $\mathfrak{q}, \bar{\mathfrak{q}}$ such that $\mathfrak{g}^{\mathbb{C}}= \mathfrak{q} \oplus \bar{\mathfrak{q}}$ and $\mathfrak{q} \cap \bar{\mathfrak{q}} = \mathfrak{h}^{\mathbb{C}}$ then it can be shown that the conditions of the Koszul theorem are met provided that  
\begin{equation} \label{eq:Jxix}
J (x) =  i x  \quad \forall x \in \mathfrak{m}^{\mathbb{C}} \cap \mathfrak{q} \:, \qquad J (\bar{x}) =  -i \bar{x} \quad \forall \bar{x} \in \mathfrak{m}^{\mathbb{C}} \cap \bar{\mathfrak{q}} \:.
\end{equation}

 \bibliography{literaturNew}
   
\bibliographystyle{JHEP}
\end{document}